\documentclass[12pt,a4paper]{article}
\usepackage{graphicx}
\usepackage{cite,./mcite}
\usepackage{xspace}
\usepackage{ifthen}
\usepackage{hyperref}
\usepackage[latin1]{inputenc}
\usepackage[dvips]{color}
  \newcommand{\ccaption}[2]{
      \caption[#1]{\small{{#2}}}
    }
    
\newcommand{\beq}{\begin{equation}}
\newcommand{\eeq}{\end{equation}}
\newcommand{\bea}{\begin{eqnarray}}
\newcommand{\eea}{\end{eqnarray}}
\newcommand{\non}{\nonumber}
\newcommand{\mc}{\mathcal}     
\newcommand{\mr}{\mathrm}

\newcommand{\be}{\begin{eqnarray*}}
\newcommand{\ee}{\end{eqnarray*}}
\newcommand{\gl}[1]{(\ref{#1})}

\begin{document}

\begin{titlepage}
\nopagebreak
  \renewcommand{\thefootnote}{\fnsymbol{footnote}}

{\begin{flushright}{
 \begin{minipage}{5cm}
   KA-TP-21-2008\\ SFB/CPP-08-88
\end{minipage}}\end{flushright}}
\vfill
\begin{center} {\Large\bf Observing Strongly Interacting Vector Boson  \\[2mm] 
Systems at the CERN Large Hadron Collider}
\end{center}
\vfill
\begin{center} {\large \bf 
C. Englert$^{1}$,
B. J\"ager$^{2,3}$, 
M. Worek$^{1,4}$, 
D. Zeppenfeld$^{1}$}
\end{center}

\vskip 0.5cm

\begin{center}
$^{1}$ ITP, Universit\"at Karlsruhe, 76128 Karlsruhe,  Germany \\ \noindent
$^{2}$ Institut f\"ur Theoretische Physik und Astrophysik, Universit\"at  
W\"urzburg, 97074~W\"urzburg, Germany
\\ \noindent
$^{3}$ KEK Theory Division, 305-0801 Tsukuba, Japan \\ \noindent
$^{4}$ Institute of Physics, University of Silesia, 40-007 Katowice, Poland \\ 
\noindent
\end{center}
\vfill
\begin{abstract}
We explore the potential of the CERN Large Hadron Collider to access a
strongly interacting electroweak symmetry breaking sector via weak boson
scattering with $W^+W^-jj$, $ZZjj$ and $W^\pm Zjj$ final states.
As examples of models with scalar or vector resonances we concentrate on
a scenario with a heavy Higgs boson and on a Warped
Higgsless Kaluza-Klein model of narrow spin-one resonances.
The signal and the most prominent background processes are evaluated using 
exact tree-level matrix elements including full off-shell and finite width 
effects for final states with two tagging jets and four leptons. 
Using double forward jet-tagging techniques, we derive dedicated cuts on
the observable jets and charged leptons to suppress Standard Model
backgrounds. We demonstrate that the LHC has substantial  
sensitivity to strong interactions in the electroweak symmetry breaking sector. 
\end{abstract}

\vskip 1cm

\today \hfill
\vfill
\end{titlepage}
%
%
%
\section{Introduction} 
An essential goal of the CERN Large Hadron Collider (LHC) 
\cite{Atlas:1999fr,Ball:2007zza} is gaining information on the mechanism which
breaks the electroweak symmetry.  Particularly promising means for probing 
electroweak symmetry breaking are provided by weak boson scattering reactions,
$VV\to VV$ (with $V$ denoting a $W^{\pm}$ or $Z$ boson). The respective 
scattering amplitudes for longitudinally polarized vector bosons 
grow with energy, thus violating unitarity beyond about 
1~TeV \cite{Veltman:1976rt,Lee:1977yc,Lee:1977eg}, when Feynman graphs with
vector bosons only are considered. Taming of this unphysical growth 
can be attained by a SM Higgs 
boson~\cite{Englert:1964et,Higgs:1964pj,Guralnik:1964eu}, but also strong 
couplings among the gauge bosons may serve to cure the
growth of the $VV$ scattering amplitudes at high 
energies~\cite{Chanowitz:1984ne,Chanowitz:1985hj,Chanowitz:1988ae}. 
Various models have been suggested in which the unitarization of these
scattering amplitudes is realized by new excitations stemming from the
compactification of extra-dimensional theories 
\cite{Csaki:2003dt,Csaki:2003zu,Agashe:2003zs}, 
based on the ideas of \cite{Randall:1999ee}.
At the LHC, weak boson scattering can be accessed via vector boson fusion 
(VBF) reactions, where the quarks emerging from the scattering protons 
emit $t$-channel weak
bosons which in turn scatter off each other. A Higgs boson as predicted by
the Standard Model (SM) would manifest itself as a relatively low mass
resonance in this reaction, but the VBF cross section would remain 
perturbatively small at di-boson masses well above the Higgs boson mass. 
In the case of
strongly interacting gauge bosons, the production rate of longitudinally
polarized gauge boson pairs $V_L V_L$ is significantly enhanced at 
$m_{VV}\approx 1$~TeV, before unitarizing effects reduce the scattering 
amplitudes. 

Signal events from strong  $V_{L}V_{L}$ scattering processes via $qq\to qq VV$
in VBF exhibit unique signatures. The decay leptons of the gauge bosons emerge
almost back-to-back in the central region of the detector with large
transverse momenta and high invariant mass. The scattered quarks give rise 
to highly
energetic jets of relatively low transverse momenta  in the forward and
backward regions. Due to the colorless weak boson exchange, the hadronic jet
activity in the central regions is very low. 
These distinctive features can be exploited to efficiently reduce background
processes with respect to the  $V_{L}V_{L}$ signals.   

The goal of this study is to refine the analyses of 
Refs.~\cite{Barger:1991ar,Barger:1993qu,Bagger:1993zf,Bagger:1995mk}
for strongly interacting electroweak symmetry breaking. Instead of using 
single forward jet tagging, as in these early analyses, we will consider
the boost invariant double forward jet-tagging techniques which have 
proven highly efficient for the search of a light Higgs boson in 
VBF~\cite{Rainwater:1997dg,Rainwater:1998kj,Rainwater:1999sd,Kauer:2000hi}. 
These more efficient jet-tagging techniques will allow us to relax 
the cuts on the $VV$ decay leptons as compared to Ref.~\cite{Bagger:1995mk}.
A second refinement is in the level of signal and background simulation.
We use parton-level 
calculations for the  processes $pp \rightarrow VVjj$ at ${\cal{O}}(\alpha^6)$ 
and ${\cal{O}}(\alpha^4 \alpha_s^2)$ including leptonic decays of 
the weak bosons, as well as a simulation of the
$t\bar{t}$, $t\bar{t}j $, and  $t\bar{t}jj$ background processes 
at ${\cal{O}}(\alpha^4
\alpha_s^2)$, ${\cal{O}}(\alpha^4 \alpha_s^3)$, and 
${\cal{O}}(\alpha^4 \alpha_s^4)$, respectively. This corresponds to tree-level
amplitudes for all processes and includes full off-shell effects for
top-quark decays $t\to Wb$ and for the leptonic decays of the weak bosons
in all signal and background processes. 
While comparable accuracy of the 
simulations has been discussed in the literature for individual reactions 
(see, e.g., Refs.~\cite{Kauer:2002sn,Accomando:2005hz,Alves:2008up}), 
an analysis of VBF signal and background processes 
with full leptonic decay correlations and off-shell effects is new.
Phenomenological studies for other production modes of extra vector
resonances 
have recently been performed in 
\cite{Birkedal:2004au,Davoudiasl:2003me,He:2007ge,Agashe:2007ki,Agashe:2008jb}. 

For the signal processes, we consider two different
models for the strongly interacting electroweak symmetry breaking.
We model unitarity conservation with a heavy and broad Standard Model scalar 
Higgs resonance, which we take as a prototype for models with strong $VV$ 
scattering. As a model with extra vector resonances, 
we adapt a Warped Higgsless scenario where unitarity violation is
postponed by the exchange of additional spin-one Kaluza-Klein (KK) resonances. 
On the basis of these two distinct examples, we show that independent 
of whether the
unitarity-restoring interactions are of a scalar or vector nature, our set of 
cuts and signal processes provides a clear signature with a large signal to 
background ratio.

Our paper is organized as follows: Section~\ref{sec:thsetup} 
outlines the theoretical setup for the two signal scenarios which we 
consider. The framework of the phenomenological 
analysis is described in Sec.~\ref{sec:frame} and here we also give 
details on the Monte Carlo calculation of the various signal and background
processes. 
In Sec.~\ref{sec:num} we present the numerical results for
expected cross sections at the LHC. 
Section~\ref{sec:summary} contains our conclusions.
%
%
\section{Theoretical Setup}
\label{sec:thsetup}
Strongly-coupled theories have a long history as extensions of the Standard
Model \cite{Farhi:1980xs,Kaplan:1983sm,Kaplan:1983fs}. In these models the
additional degrees of freedom, needed to unitarize longitudinal gauge boson
scattering, originate from a strongly-interacting sector that produces scalar
and vectorial composites. While electroweak symmetry breaking is stabilized at
the electroweak scale, the compositness scale is in principle a free
parameter. The unitarizing mass spectrum can be rather heavy and broad, due to 
strong couplings in the composite sector. As such models are intrinsically 
non-perturbative, there are large theoretical uncertainties on the theory's 
parameters that can only roughly be estimated by Naive Dimensional 
Analysis  \cite{Manohar:1983md}. This poses a huge challenge for modelling 
LHC phenomenology. For this reason we use unitarity of longitudinal $VV$
scattering at the TeV scale as key ingredient to model the strongly interacting sector, 
focusing on two 
distinct scenarios: In the first one, we adapt a heavy and broad scalar 
resonance, while in the second one, longitudinal gauge boson scattering 
is unitarized by vectorial resonances in the Warped Higgsless Model of 
Refs.~\cite{Csaki:2003dt,Csaki:2003zu}.

\subsection{Scalar Resonance}
\label{sec:heavyh}

Within the SM, unitarization of longitudinal VV scattering is achieved by 
adding the contributions of a scalar resonance of zero isospin, the Higgs 
boson, to the gauge boson exchange graphs which are mandated by the gauge 
symmetry. Working within the SM, precision data, in particular the results 
of LEP and SLC on various four-fermion processes combined with the direct 
Higgs search at LEP, constrain the mass of the SM Higgs boson to lie 
inside the 100 to 200 GeV region~\cite{Barate:2003sz,Erler:2008ek}. Strictly 
within the SM, a heavy scalar resonance, with a mass of order 1~TeV, is 
ruled out as a model for unitarized weak boson scattering.

These SM Higgs boson mass bounds might be misleading, however, if other 
new physics contributions to four-fermion amplitudes (partially) cancel 
the virtual contributions of a heavy Higgs boson to the $S$ and $T$ 
parameters~\cite{Peskin:1990zt,Peskin:1991sw,Barbieri:2004qk,Peskin:2001rw}, thus mitigating the 
constraints from precision experiments. Since the precision observables, 
with their strong focus on four fermion-amplitudes, and weak boson 
scattering amplitudes are independent entities in sufficiently general 
models of new physics, we ignore the constraints from precision data in 
the following and consider, as a phenomenological model, unitarization of 
weak boson scattering by a scalar resonance with quantum numbers and 
couplings identical to a heavy SM Higgs boson. We include $s$-, $t$- 
and/or $u$-channel channel exchange of this resonance and use $m_H = 
1$~TeV and a fixed width $\Gamma_H=0.5$~TeV as a toy-model
for demonstration purposes. 
This fixed width is included for time-like and space-like propagators, in 
analogy to the complex mass scheme for the gauge boson propagators. 
More general model parameters 
of a heavy scalar resonance can easily be implemented in the 
\textsc{Vbfnlo} program~\cite{VBFNLO} which we use 
for all signal simulations.

\subsection{Vectorial Resonances}
\label{sec:kk}
As an example of unitarization with vectorial resonances we consider a
phenomenological version of the Warped Higgsless model of Refs.~\cite{KK1,Englert:2008wp}. 
Using the AdS/CFT correspondence
\cite{Maldacena:1997re,ArkaniHamed:2000ds,Rattazzi:2000hs}, the Warped Higgsless scenario 
can be considered as a particular type of a strongly-interacting 
Walking Technicolor theory
\cite{ArkaniHamed:2000ds}, yet being calculable by perturbative means from
a bulk-gauged effective theory defined on a slice of a five dimensional Anti-de Sitter space. 
In these scenarios the growth of the amplitude in longitudinal gauge boson
scattering is tamed by the exchange of heavy spin-1 Kaluza-Klein excitations, 
$W^\pm_k$ and $Z_k$. 
Demanding sum rules for the quartic and triple vector boson 
couplings \cite{Birkedal:2004au, SekharChivukula:2008mj}, 
\bea
\label{sumww1}
g_{W_1W_1W_1W_1}&=& \sum_{k\geq 0} g_{W_1W_1Z_k}^2\,,\\
\label{sumww2}
4m_{W_1}^2g_{W_1W_1W_1W_1} &=& 3\sum_{k\geq 1} m_{Z_k}^2 g_{W_1W_1Z_k}^2\,, \\
\label{sumwz1}
g_{W_1W_1Z_1Z_1}&=&\sum_{k\geq 1} g_{W_kW_1Z_1}^2\,,\\
\label{sumwz2}
2(m_{Z_1}^2+m_{W_1}^2)g_{W_1W_1Z_1Z_1}&=& \sum_{k\geq 1} g_{W_kW_1Z_1}^2 
\left( 3m_{W_k}^2 - {(m_{Z_1}^2-m_{W_1}^2)^2\over m_{W_k}^2 }\right) \,,
\eea
results in good high energy behaviour of the $V_LV_L$ scattering amplitude. In \gl{sumww1}-\gl{sumwz2}
$k$ labels the Kaluza-Klein states, and $k=0,1$ identifies 
the massless and massive gauge bosons
of the SM, respectively. 
We focus on a scenario where the new additional massive vector bosons 
have vanishing couplings to SM-fermions, and include states up to $W_4$~and~$Z_6$. 

Higgsless symmetry breaking has already been studied in various realizations \cite{Csaki:2003dt,Csaki:2003zu,Cacciapaglia:2004rb,SekharChivukula:2005xm,Agashe:2003zs,Birkedal:2004au,ArkaniHamed:2001nc,He:2007ge,Davoudiasl:2003me,Agashe:2007ki,Agashe:2008jb}. In this paper we do not attempt to 
construct a realistic model of Higgsless symmetry breaking, but we solely use the quoted sum rules as a phenomenological paradigm of unitarization with iso-vectorial resonances. For a more detailed discussion on the implementation of the sum rules and the KK mass spectrum, we refer 
the reader to a separate publication \cite{Englert:2008wp}. 
%
\section{Framework of the Analysis}
\label{sec:frame} 
Throughout this study, we consider vector boson pair production  
in association
with two tagging jets, 
$
pp \to VVjj
$, 
with subsequent leptonic decays of the gauge bosons 
in proton-proton collisions at the LHC  
with  a center-of-mass energy
of 14~TeV. 
If strong interactions among longitudinally polarized vector bosons 
are realized in nature, $V_L V_L\to V_L V_L$ scattering is expected to be 
enhanced at large invariant mass. 
In contrast, the scattering of transversely polarized gauge bosons 
$V_T$ is dominated by the same weak gauge interactions as in the 
SM light Higgs boson scenario and, thus, remains perturbative 
throughout the entire $VV$ invariant mass range. The $V_T V_T \to V_T V_T$ 
and $V_L V_T \to V_L V_T$ contributions to vector boson scattering 
must be considered as an irreducible background to the signature of strong
gauge boson interactions, which we wish to isolate. 
We thus define the VBF 
``signal'' in EW $pp\to VVjj$ production as the enhancement of the
cross section over the  SM prediction with a light Higgs boson. In the 
heavy Higgs boson scenario this is 
\begin{equation}
\label{eq:signal-hh}
\sigma_{S} \equiv \sigma_{SM}(m_H=1 ~\textnormal{TeV}) - 
\sigma_{SM}(m_H=100 ~\textnormal{GeV}).
\end{equation}
As an alternative realization of electroweak symmetry breaking we consider 
the Warped 
Higgsless Kaluza-Klein model described in Sec.~\ref{sec:kk}. In this context 
we define  
\begin{equation}
\label{eq:signal-kk}
\sigma_{S} \equiv \sigma_{KK} - 
\sigma_{SM}(m_H=100 ~\textnormal{GeV}).
\end{equation}

Backgrounds arise from  QCD-induced and non-resonant electroweak (EW)
reactions with the same final-state configuration as the signal, 
at $\mc{O}(\alpha^4\alpha_s^2)$ and $\mc{O}(\alpha^6)$, respectively. 
For the $W^+W^-jj$ channel, the production processes 
$t\bar{t}$, $t\bar{t}j$ and $t\bar{t}jj$ 
at ${\cal{O}}(\alpha^4 \alpha_s^2)$,
${\cal{O}}(\alpha^4 \alpha_s^3)$, and
${\cal{O}}(\alpha^4 \alpha_s^4)$, 
respectively, have to be considered as copious background sources also. 
Via their decay chains,
the $t\bar t$ pairs give rise to the same combination of charged leptons in the
final state as the VBF signal process. 

Since the principle subject of this study is the investigation of strongly
interacting gauge boson systems, we do not consider signal processes deriving 
from Yukawa couplings of the Higgs boson to fermions, 
such as gluon-induced $Hjj$ production. 

 
\subsection{Details of the Calculation}
The calculation of cross sections and kinematic distributions for all 
signal and background processes introduced above is performed with two
independent computer programs featuring full tree-level matrix elements: 
\begin{itemize}
\item
Results for all but 
Kaluza-Klein signal reactions are generated with \textsc{Helac-Phegas}, a
completely automatic Monte-Carlo event generator 
\cite{Kanaki:2000ey,Papadopoulos:2000tt,Papadopoulos:2005ky,
Cafarella:2007pc,HELAC}, which calculates matrix elements through
Dyson-Schwinger off-shell recursive equations. The package  
provides events for arbitrary 
parton-level processes in the most 
recent Les Houches Accord format \cite{Alwall:2006yp} and 
has successfully been tested for scattering reactions at a
future linear collider \cite{Gleisberg:2003bi} 
and at the LHC \cite{Alwall:2007fs}.  
\item
The EW VBF and Kaluza-Klein signal and background processes are tackled with 
the tree-level version of \textsc{Vbfnlo} \cite{VBFNLO}, a parton-level 
Monte-Carlo program for VBF-type reactions. 
For the QCD $VVjj$ processes, we implemented  
\textsc{MadGraph}-generated amplitudes \cite{Maltoni:2002qb,Alwall:2007st}
into the framework of \textsc{Vbfnlo}. 
Results for the $t\bar{t}$, $t\bar{t}j$, and $t\bar{t}jj$ reactions are
generated with the codes of Ref.~\cite{Kauer:2002sn}. 
\end{itemize}
Making sure that the different programs yield the same results 
provides an excellent check of our calculation. 
In particular, \textsc{Helac-Phegas} agrees at least at the level of 1\% with 
the top-backgrounds of  Ref.~\cite{Kauer:2002sn}, and even better with 
the \textsc{MadGraph}-type implementation of the QCD $VVjj$ backgrounds 
in  \textsc{Vbfnlo}, irrespectively of the cuts applied. 

For our numerical studies we use the 
CTEQ6L1 parton distribution functions 
\cite{Pumplin:2002vw,Stump:2003yu} at leading order (LO) 
with $\alpha_s(M_Z) = 0.130$. We have chosen $M_Z = 91.188$~GeV, $M_W = 
80.423$~GeV,
and $G_F = 1.166\times 10^{-5}/$~GeV$^2$ as electroweak input parameters.
The other parameters, $\alpha$ and $\sin^2\theta_W$, 
are computed thereof via LO EW relations. The masses of the top and 
bottom quarks are set to $m_t = 172.5$~GeV and $m_b = 4.4$~GeV, respectively. 
Contributions from $b$- and $t$-quarks in 
the initial state are neglected throughout. 
In \textsc{Helac-Phegas}, finite width effects in massive vector boson and top
 quark
propagators are taken into account by 
the complex mass scheme of 
Refs.~\cite{Denner:1999gp,Denner:2005fg,Denner:2006ic}. 
Both, in \textsc{Vbfnlo} and in the code of Ref.~\cite{Kauer:2002sn},  
unstable particles are treated via modified versions 
\cite{Oleari:2003tc,Kauer:2001sp} of the complex mass scheme.
Spin and color correlations of the final state particles are
taken into account without any approximations.
Final state partons are recombined into jets according to the $k_T$
algorithm~\cite{Catani:1992zp,Catani:1993hr,Ellis:1993tq} with resolution 
parameter 0.7. 
In the following, we outline the process-specific settings of our analysis. 


\subsubsection{EW $VVjj$ production}
%
\begin{figure}[!t]
\begin{center}
\includegraphics[angle=-90,width=1.0\textwidth]{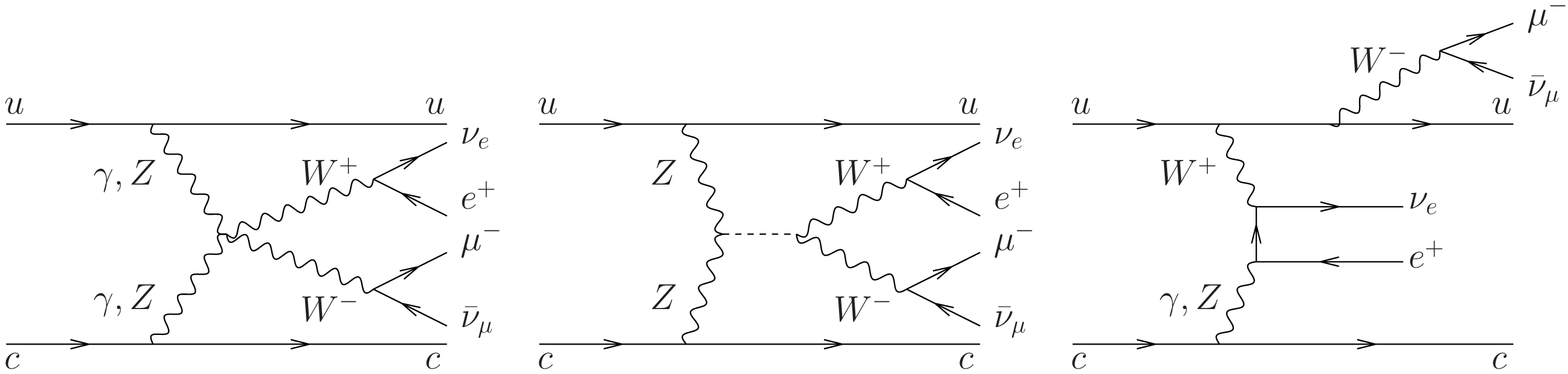}
\end{center}
\vskip -0.4cm
\ccaption{}{\label{fd-vv}\it 
Examples of Feynman-graph topologies contributing to 
EW~$W^+W^-jj$ production at $\mc{O}(\alpha^6)$.}
\vskip 0.4cm
\end{figure}
%
EW $VVjj$ production 
mainly proceeds via the fusion of weak bosons in the $t$-channel in
quark-(anti)quark scattering processes like $qq'\to qq'VV$.
In experiment, however, leptons rather than vector bosons are identified. We
therefore focus on the reactions 
\bea
pp&\to& \,\ell^+\nu_\ell \,\ell'^-\bar\nu_{\ell'}\,jj,\non\\
pp&\to& \,\ell^+\ell^- \,\ell'^+\ell'^-\,jj\,,\non\\
pp&\to& \,\ell^+\ell^- \,\nu_{\ell'} \bar\nu_{\ell'}\,jj\,,\\
pp&\to& \,\ell^+\nu_\ell \,\ell'^+\ell'^-\,jj,\non\\
pp&\to& \,\ell^-\bar\nu_\ell \,\ell'^+\ell'^-\,jj\,\non
\eea
at $\mc{O}(\alpha^6)$, which include the resonant $VVjj$ production processes 
with subsequent leptonic decays and additional single- and non-resonant 
diagrams, see Fig. \ref{fd-vv}. 
We only simulate decays of the weak 
bosons to different lepton generations, e.g. 
$W^+W^-\rightarrow e^+\nu_e\mu^-\bar\nu_\mu$. 
Same-generation lepton interference effects as occurring in
$W^+W^-\rightarrow e^+\nu_e e^-\bar\nu_e$ are neglected for all production
channels. However, we adjust counting factors to correspond to the production 
of all combinations of charged leptons of the first two generations.
In case of $Z \rightarrow \nu_\ell \bar\nu_{\ell}$ we sum over three neutrino 
generations, i.e. $\nu_\ell\bar{\nu}_\ell=\nu_e\bar{\nu}_e, 
\nu_\mu\bar{\nu}_\mu, \nu_\tau \bar{\nu}_\tau$.
For brevity, we will refer to these reactions as EW  $W^+W^-\,jj$,
$ZZ\,jj\rightarrow 4\ell\,jj$, $ZZ\,jj\rightarrow 2\ell 2\nu\,jj$, $W^+Z\,jj$, 
and $W^-Z\,jj$ production, respectively, even though we are always 
considering  leptonic final states.

As discussed in 
Refs.~\cite{Jager:2006zc,Jager:2006cp,Bozzi:2007ur,Ciccolini:2007ec}, 
compared to the dominant $t$-channel configurations, 
contributions from $s$-channel electroweak 
boson exchange and identical fermion effects are
negligible in the phase-space regions where $VVjj$ production is observed 
experimentally.
They are therefore disregarded for our analysis. In Ref.~\cite{Bozzi:2007ur} 
it has been demonstrated that next-to-leading order (NLO) QCD effects can 
be well approximated also in distributions by a proper choice of 
the factorization scale, $\mu_F$, in the LO calculation: for each
fermion line choose the momentum transfer $Q$  between the
respective initial- and final-state quarks. We therefore set 
$\mu_F=Q$ for all EW $VVjj$ processes.


\begin{figure}[!t]
\begin{center}
\includegraphics[width=0.4\textwidth]{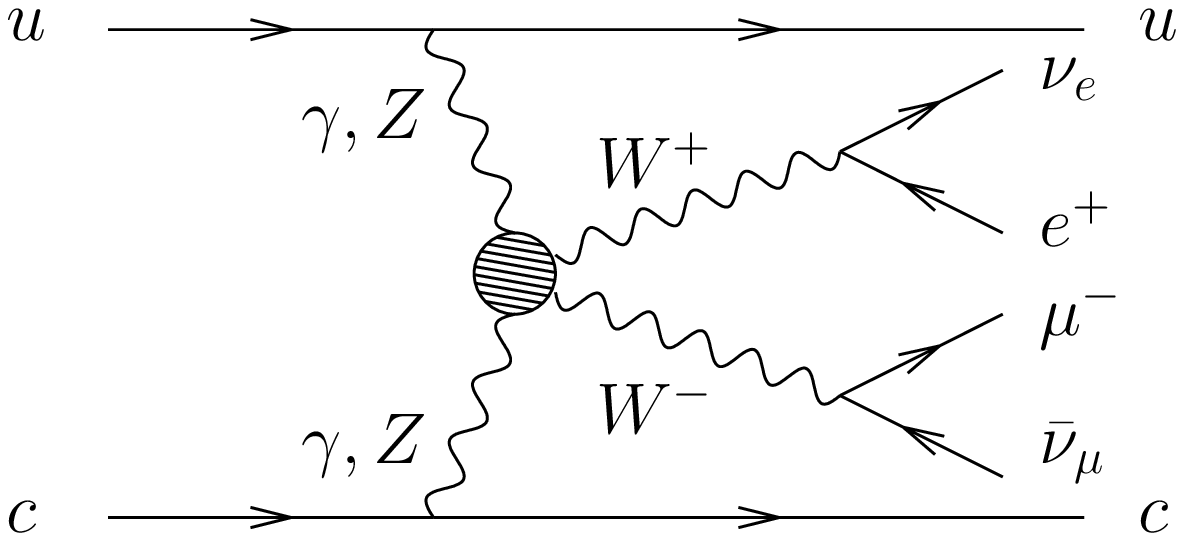}
\end{center}
\vskip -0.4cm
\ccaption{}{\label{fig:wbftop}\it 
Modified weak boson fusion topology. The shaded area contains different 
Kaluza-Klein intermediate states.}
\vskip 0.4cm
\end{figure}
%
\subsubsection{Higgsless $VVjj$ production}
The implementation of the Kaluza-Klein scenario described in 
Sec.~\ref{sec:kk} into the \textsc{Vbfnlo} framework is described in detail in
 Refs.~\cite{KK1,Englert:2008wp}. The leptonic tensors for 
subamplitudes such as $ZZ\to e^+\nu_e\mu^-\bar\nu_\mu$ in 
Fig.~\ref{fig:wbftop} have been extended by the different Kaluza-Klein 
intermediate states. 
We discard interactions of non-SM Kaluza-Klein gauge bosons with the 
light SM fermions. 
The coupling of the Kaluza-Klein $W_{k}$ and $Z_{k}$ to the $W_1$ 
and $Z_1$ steeply drops off
with the Kaluza-Klein index $k$. For $k \ge 3$, contributions of Kaluza-Klein
excitations to cross sections and distributions are tiny \cite{Englert:2008wp}.
In our studies 
we include all Kaluza-Klein states up to $W_{4}$ and $Z_{6}$ 
with the masses and widths as given in Tab.~\ref{tab:kk}.

Gauge boson pair production in the presence of Kaluza-Klein excitations 
proceeds analogously to Higgs-mediated $VVjj$ production. In order to absorb 
the dominant NLO-QCD effects we therefore use the same factorization scales
as for EW $VVjj$ production, i.e.\  $\mu_F=Q$~\cite{KK1,Englert:2008wp}.
%
\begin{table}[t]
\begin{center}
\begin{tabular}{c|c|c}
\hline
&&\\
Particle &  $m$ ~[GeV] &  $\Gamma$ ~[GeV] \\
&&\\
\hline
&&\\
 $W_{2}$ & 700  & 13.7\\
 $Z_{2}$ & 695  & 8.7 \\
 $Z_{3}$ & 718 & 6.4\\
&&\\
\hline
&&\\
$W_{3}$ & 1106 & 31.0  \\
$Z_{4}$ & 1112 & 26.5 \\
&&\\
\hline
&&\\
$W_{4}$ & 1585 & 56.5 \\
$Z_{5}$ & 1580 & 31.7 \\
$Z_{6}$ & 1605 & 24.1 \\
&&\\
\hline
\end{tabular}
\ccaption{}{\label{tab:kk} \it Masses and widths of the Kaluza-Klein 
resonances used in the simulation. The spectrum corresponds to a Planck brane localization 
$R=9.75\times 10^{-9}~\mathrm{GeV}^{-1}$.}
\end{center}
\end{table}
%
%
\subsubsection{QCD $VVjj$ production}
QCD-induced $VVjj$ production calculated at order
${\cal{O}}(\alpha^2\alpha_s^2)$ includes the production processes 
\bea
qq\to qq VV\,,\quad qg\to qgVV\,,
\eea
with subsequent leptonic decays, and all crossing-related reactions. 

For these processes we use
$\mu_F =\min (p_{T j_1},p_{T j_2})$. 
The renormalization scale is chosen such that the strong coupling  
factor takes the form 
$\alpha_{s}^2=\alpha_{s}(p_{T j_1})\cdot\alpha_{s}(p_{T j_2})$, i.e.\ 
the transverse momentum of each parton is taken as the 
relevant scale for its production. 
%
%
\subsubsection{$t\bar t+\mr{jets}$ production}
Due to the large top quark production rate at the LHC and because the branching
ratio $B(t\to Wb)$ is essentially 100\%, $t\bar t +\mr{jets}$ processes
constitute a major background to EW~$W^+W^-jj$ production. 
We consider the reactions $pp\to t\bar t$, $t\bar tj$, and $t\bar tjj$ 
which include full off-shell and finite width top and $W$ effects and take into
account the double-resonant, single-resonant and non-resonant contributions
at order ${\cal{O}}(\alpha^2\alpha_s^2)$, ${\cal{O}}(\alpha^2\alpha_s^3)$
and ${\cal{O}}(\alpha^2\alpha_s^4)$, respectively. To avoid double counting, 
the top-quark backgrounds are separated into three categories, depending on
whether two, one or zero $b(\bar{b})$  quarks are identified as tagging jets 
and are referred to as $t\bar{t}$, $t\bar{t}j$ and $t\bar{t}jj$ background,
respectively.  
When combining these processes, we proceed as follows: 
For $t\bar{t}jj$ production both tagging jets are required to arise 
from massless partons, while in the $t\bar{t}j$
case exactly one tagging jet is allowed to emerge from a $b$ or $\bar b$ 
quark. For $t\bar{t}$ production both tagging jets stem from $b$
quarks \cite{Rainwater:1999sd,Kauer:2000hi}.
When presenting cross sections and kinematic distributions, the 
three $t\bar{t}+\mr{jets}$
backgrounds are combined for 
clarity even though their individual distributions are slightly
different. 

In all cases, the factorization scale is chosen as $\mu_F = \min(m_{T_i})$ 
of the top quarks and additional jets, where each $m_{T_i}$ is given by the
transverse momentum and mass of the respective entity $i$ as
\beq
m_{T_i} = \sqrt{p_{T_i}^2+m_i^2}\,.
\eeq
The overall strong coupling factors for the $t\bar t + n \,\mr{jets}$ cross
section are calculated as 
$
(\alpha_s)^{n+2} = \prod_{i=1}^{n+2}\alpha_s(m_{T_i})
$.
%
%
\subsection{Selection Cuts}
In order to suppress the backgrounds with respect to the 
signal processes, the design of dedicated selection cuts is essential. 
For our analysis we have developed various sets of cuts, which are given as
follows:
\begin{itemize}
\item[I.]\textsc{Inclusive cuts:}
Basic selection cuts need to be introduced to render  
our calculation of the production cross sections
of all signal and background processes finite. 
This is achieved by identifying all final state massless partons with
high transverse momentum jets. The two jets of largest transverse 
momentum are called ``tagging jets'' and are required to carry
\begin{equation}
\label{cuts:inc1}
p_{Tj}^\mr{tag} > 30 ~\textnormal{GeV}\,.
\end{equation}
All jets need to lie in the rapidity-range accessible to the detector, 
\begin{equation}
|\eta_{j}| < 4.5\,,
\end{equation}
and are supposed to be well-separated,
\begin{equation}
~~\Delta R_{jj} =\sqrt{(\eta_{j_{1}}-\eta_{j_{2}})^2
+(\phi_{j_{1}}-\phi_{j_{2}})^2}  > 0.7\,,
\end{equation}
with $\eta_j$ denoting the jet rapidity and 
$\Delta R_{jj}$ the separation of any pair of jets in the
rapidity-azimuthal angle plane. 
For all $VVjj$ production processes, the tagging
jets are identified with the massless final-state partons of the reaction. For
the $t\bar t + \mr{jets}$ backgrounds, the tagging jets can stem from a 
massless quark or
gluon, or from the decay products of the top quarks.  

In order to ensure well-observable isolated charged leptons in the
central-rapidity region, we require
\begin{equation}
p_{T\ell} > 20 ~\textnormal{GeV},
~~|\eta_\ell| < 2.5, ~~\Delta R_{\ell j} > 0.4\,,
\end{equation}
where $\Delta R_{\ell j}$ stands for the separation of a charged
lepton from any jet. Since any $b$-quark close to a charged lepton is
very likely to also spoil lepton isolation, we require
$\Delta R_{\ell b} > 0.4$ even if the $b$-quark is too soft to qualify
as a jet.
Finally, a  cut on the invariant mass  $m_{\ell\ell}$ of
two charged leptons of the same flavor is applied to avoid 
virtual photon singularities stemming from quasi-collinear
$\gamma^{*}\rightarrow \ell^+\ell^-$ decays,
\begin{equation}
m_{\ell\ell} > 15 ~\textnormal{GeV}. 
\end{equation}
\item[II.]\textsc{VBF cuts:}
VBF events are characterized by two tagging jets in the far 
forward and backward
regions of the detector, while the leptonic decay products of the vector
bosons are typically located in the central-rapidity range between the jets. 
To favor such configurations, we demand that the charged leptons fall 
between the tagging jets, 
\begin{equation}
\eta_{j,min}^{tag} < \eta_\ell < \eta_{j,max}^{tag} \,,
\end{equation} 
which are well-separated in rapidity,
\begin{equation}
\Delta \eta_{jj}
     =|\eta_{j_1}^{tag}-\eta_{j_2}^{tag}| > 4\,,
\end{equation}
and 
occupy opposite detector hemispheres, 
\begin{equation}
\label{cuts:vbf3}
\eta_{j_1}^{tag} \times \eta_{j_2}^{tag} < 0\,.
\end{equation}
Furthermore, the tagging jets are required to have a large invariant mass, 
\begin{equation}
\label{cuts:vbf4}
m_{jj} > m_{jj}^{min}\,,
\end{equation}
where $m_{jj}^{min} = 1000$~GeV for the $W^+W^-jj$ signal and background 
processes
and  $m_{jj}^{min} = 500$~GeV for all other channels. 

To illustrate the significance of the $m_{jj}$ cut, the invariant mass 
distribution of the two tagging jets in $pp \rightarrow W^+W^-jj$ is 
shown in Fig.~\ref{ww1-inter}, after applying the cuts of 
Eqs.~(\ref{cuts:inc1})-(\ref{cuts:vbf3}) and requiring 
$p_{T}(\ell) > 100$~GeV. For reducing the $t\bar t+\mr{jets}$ backgrounds, 
additionally a $b$-veto and a central jet veto have been imposed, as 
discussed below. While large invariant masses of the tagging jets are 
characteristic for VBF processes, QCD-induced reactions tend to peak 
at small values of $m_{jj}$. Requiring $m_{jj} > 1000$~GeV thus 
efficiently suppresses contributions from $t\bar t+\mr{jets}$ and 
QCD $VVjj$ production with respect to the signal processes. 
%
\begin{figure}
\begin{center}
\includegraphics[width=7.5cm,height=7.5cm]{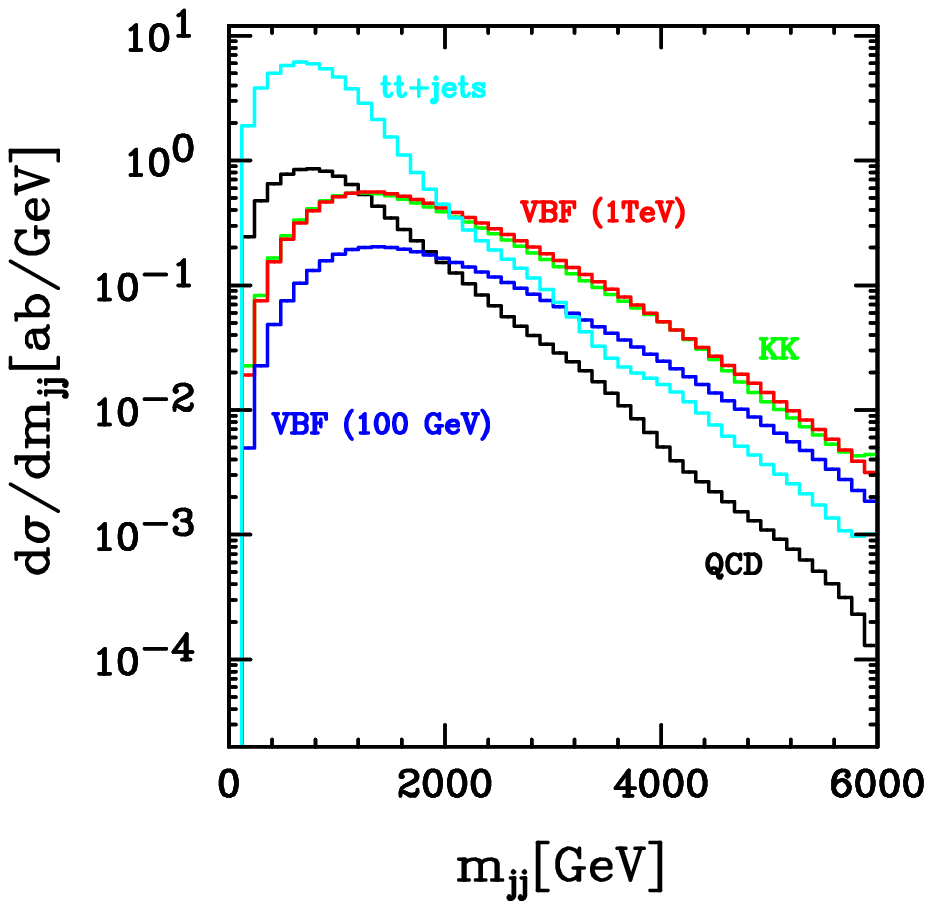}
\end{center}
\vskip -0.4cm
\ccaption{}{\label{ww1-inter} \it
Invariant mass distribution of the two tagging jets for $pp \rightarrow W^+W^-
jj$ after imposing the cuts of Eqs.~(\ref{cuts:inc1})-(\ref{cuts:vbf3}), 
a $b$-veto, a CJV, 
and requiring $p_{T}(\ell) > 100$~GeV. Plotted are results for the 
heavy Higgs boson scenario, the Higgsless Kaluza-Klein model, and the
relevant SM backgrounds. }
\vskip 0.4cm
\end{figure}
%
\item[III.]\textsc{Leptonic cuts:}
In all channels, the signal processes feature energetic 
leptons of high $p_T$ and large invariant mass. The
decay products of the backgrounds are less back-to-back in the transverse plane
and are characterized by lower transverse momenta. 
These features suggest the application of extra selection cuts 
specific to each decay channel:
\begin{itemize}
\item[$\bullet$]{$ZZ jj\to 4\ell\, jj$:
\bea
&&m_{ZZ} >  500 ~\textnormal{GeV}\,,
\non\\
&&p_T(\ell\ell) > 0.2 \times m_{ZZ}\,.
\eea
Here, $m_{ZZ}$ is the invariant mass of the four-lepton system, 
and $p_T(\ell\ell)$ the
transverse momentum of two same-flavor charged leptons. 
}
\item[$\bullet$]{$ZZ jj \rightarrow 2\ell 2\nu\,jj$:
\bea
&&m_T(ZZ) > 500 ~\textnormal{GeV}\,,
\non\\
&&p_{T}^{miss} > 200 ~\textnormal{GeV}\,,
\eea
with $p_{T}^{miss}$ being the transverse momentum of the neutrino system and 
\beq
m^{2}_{T}(ZZ)=[ \sqrt{m_{Z}^{2}+p_{T}^{2}(\ell\ell)} + 
\sqrt{m_{Z}^{2}+(p_{T}^{miss})^2} ]^2-
[ \vec{p}_{T}(\ell\ell) +\vec{p}_{T}^{\;miss}]^2\,.
\eeq
}
\item[$\bullet$]{
$W^\pm Z jj$: 
\bea
&&m_T(WZ) > 500 ~\textnormal{GeV}\,,
\non\\
&&p_T^{miss} > 30 ~\textnormal{GeV}\,,
\eea
where 
\beq
m^{2}_{T}(WZ)=[ \sqrt{m^{2}(\ell\ell\ell)+p_{T}^{2}(\ell\ell\ell)} +
  |p_{T}^{miss}| ]^2 - [ \vec{p}_{T}(\ell\ell\ell) +\vec{p}_{T}^{\;miss} ]^2\, ,
\eeq
with $m(\ell\ell\ell)$ and $p_{T}(\ell\ell\ell)$ denoting the invariant mass and
transverse momentum of the charged-lepton system, respectively. 
}
\item[$\bullet$]{$W^+W^- jj$:
\bea
\label{eq:ww_lep_cuts}
&&p_{T\ell}>100 ~\textnormal{GeV} \,,
\non\\
&&\Delta p_{T}(\ell\ell)=
|\vec{p}_{T,\ell_1}-\vec{p}_{T,\ell_2}|>250~\textnormal{GeV} \,,
\non\\
&&m_{\ell\ell} > 200 ~\textnormal{GeV} \,,
\non\\
&&\min\,(m_{\ell j}) > 180 ~\textnormal{GeV} \,,
\eea
where $\Delta p_{T}(\ell\ell)$ is the difference between the transverse momenta
of the two charged decay leptons, and $\min(m_{\ell j})$ the minimum invariant 
mass of a  tagging jet and any charged lepton. 
} 
\end{itemize}

To motivate this set of selection cuts, we show representative distributions 
for the $pp \rightarrow W^+W^-jj$ channel in the following.
In Fig.~\ref{ww2-inter}, 
%
\begin{figure}
\begin{center}
\includegraphics[width=7.5cm,height=7.5cm]{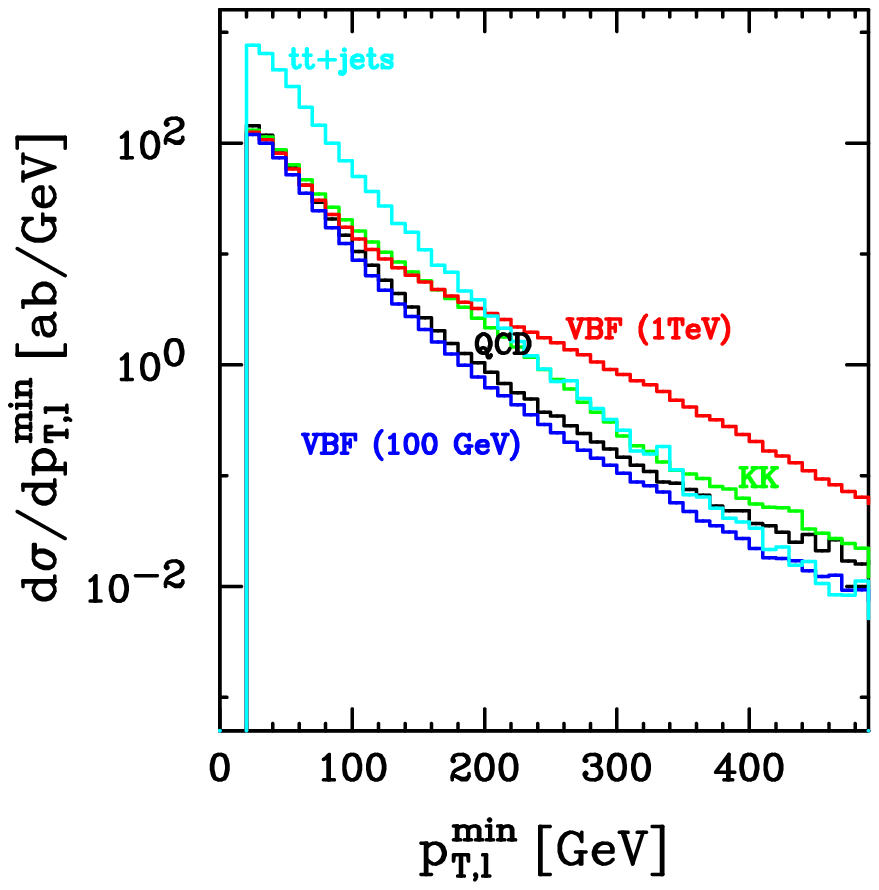}
\end{center}
\vskip -0.4cm
\ccaption{}{\label{ww2-inter}\it
Transverse momentum distribution of the softest charged lepton for 
$pp \rightarrow W^+W^-jj$ 
after 
imposing  the cuts of Eqs.~(\ref{cuts:inc1})-(\ref{cuts:vbf4}), 
a $b$-veto, and a CJV. Plotted are results for the 
heavy Higgs boson scenario, the Higgsless Kaluza-Klein model, and the
relevant SM backgrounds.}
\vskip 0.4cm
\end{figure}
%
the transverse momentum distribution of the softest charged lepton is shown 
after imposing  the cuts of Eqs.~(\ref{cuts:inc1})-(\ref{cuts:vbf4}), a 
$b$-veto, and a central jet veto. While the heavy-Higgs and the Kaluza-Klein 
distributions can barely be distinguished from the QCD and EW backgrounds at 
low transverse momenta, the signal cross sections start to deviate from the 
EW $WWjj$ background at about $p_{T\ell}\approx 100$~GeV. Removing events 
with $p_{T\ell}< 100$~GeV therefore helps to suppress irreducible backgrounds 
from SM-like $W^+W^-jj$ production processes. For reducing the still sizeable 
$t\bar t+\mr{jets}$ cross sections, additional cuts are necessary. 

Figure~\ref{ww4-inter}~(a)    
%
\begin{figure}
\begin{center}
\includegraphics[width=7.5cm,height=7.5cm]{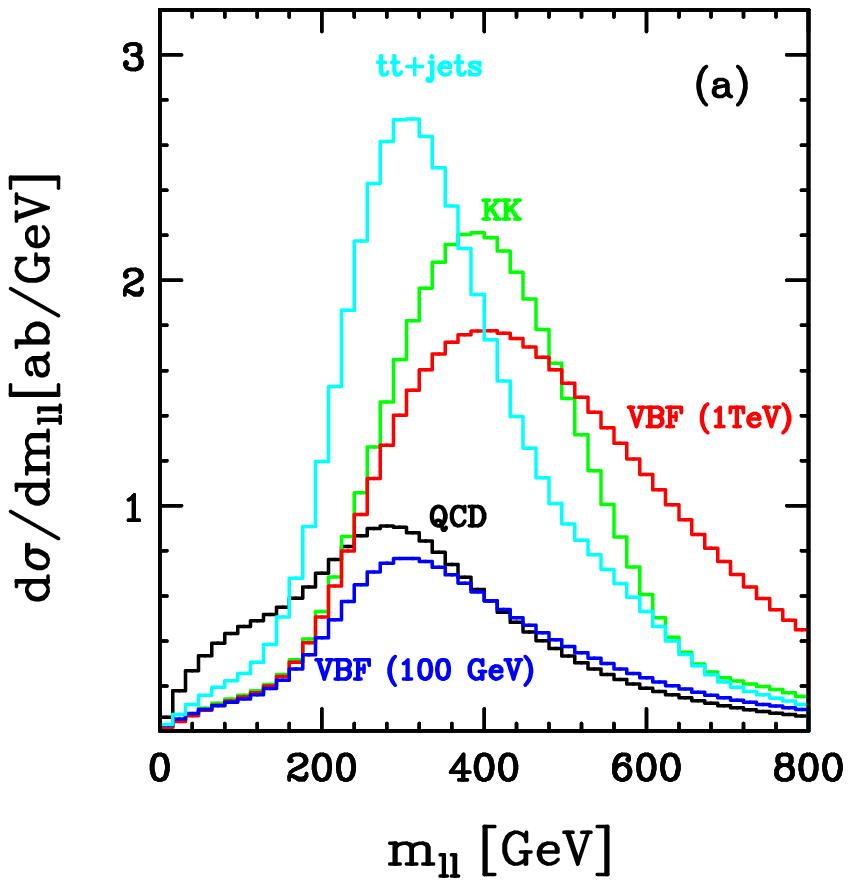}
\includegraphics[width=7.5cm,height=7.5cm]{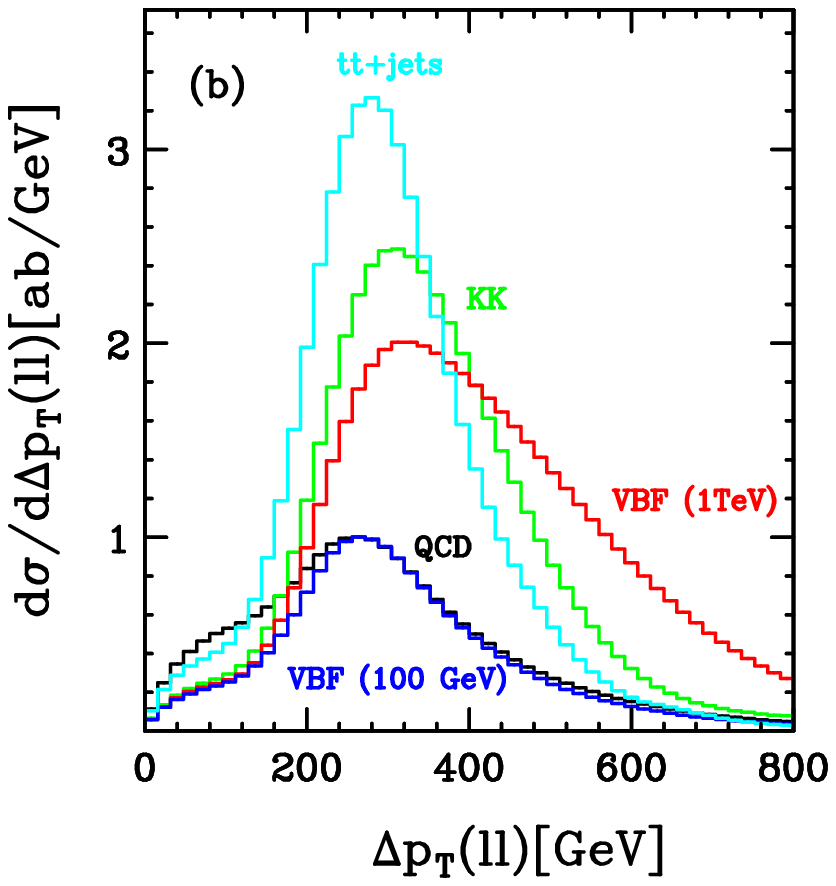}
\end{center}
\vskip -0.4cm
\ccaption{}{\label{ww4-inter} \it
Invariant mass distribution of the two charged leptons (a) and 
difference between their transverse momenta (b) for the 
$pp \rightarrow W^+W^-jj$ process after imposing 
the cuts of Eqs.~(\ref{cuts:inc1})-(\ref{cuts:vbf4}), 
a $b$-veto, a CJV, 
and requiring $p_{T}(\ell) > 100$~GeV and $\min(m_{lj}) > 180$~GeV.
Plotted are results for the 
heavy Higgs boson scenario, the Higgsless Kaluza-Klein model, and the
relevant SM backgrounds.}
\vskip 0.4cm
\end{figure}
%
displays the invariant mass distribution of the two charged final-state 
leptons after all inclusive and VBF cuts have been applied,  and a $b$-veto, 
a central jet veto, $p_{T}(\ell) > 100$~GeV and $\min(m_{lj}) > 180$~GeV have
 been imposed. In the heavy-Higgs and Kaluza-Klein signal processes, the 
invariant mass distribution peaks at rather large values of $m_{\ell\ell}$, 
while smaller invariant masses are preferred by the background processes, 
which therefore can be reduced considerably by requiring 
$m_{\ell\ell} > 200$~GeV.  

Choosing the cut on the difference in the transverse momenta of the decay 
leptons is a subtle issue, as the peaks of the signal and background 
distributions are located rather closely in $\Delta p_{T}(\ell\ell)$, 
see Fig.~\ref{ww4-inter}~(b). Selecting events with 
$\Delta p_{T}(\ell\ell)>250~\textnormal{GeV}$ turns out to be a reasonable
 choice, however, which suppresses contributions from the 
$t\bar{t} + \mr{jets}$ and the QCD $VVjj$ processes, while the 
Kaluza-Klein and heavy-Higgs cross sections are retained to a large extent. 

The cut on the minimum invariant mass of the tagging jet and any 
charged lepton, depicted in Fig.~\ref{ww3-inter},  
%
\begin{figure}
\begin{center}
\includegraphics[width=7.5cm,height=7.5cm]{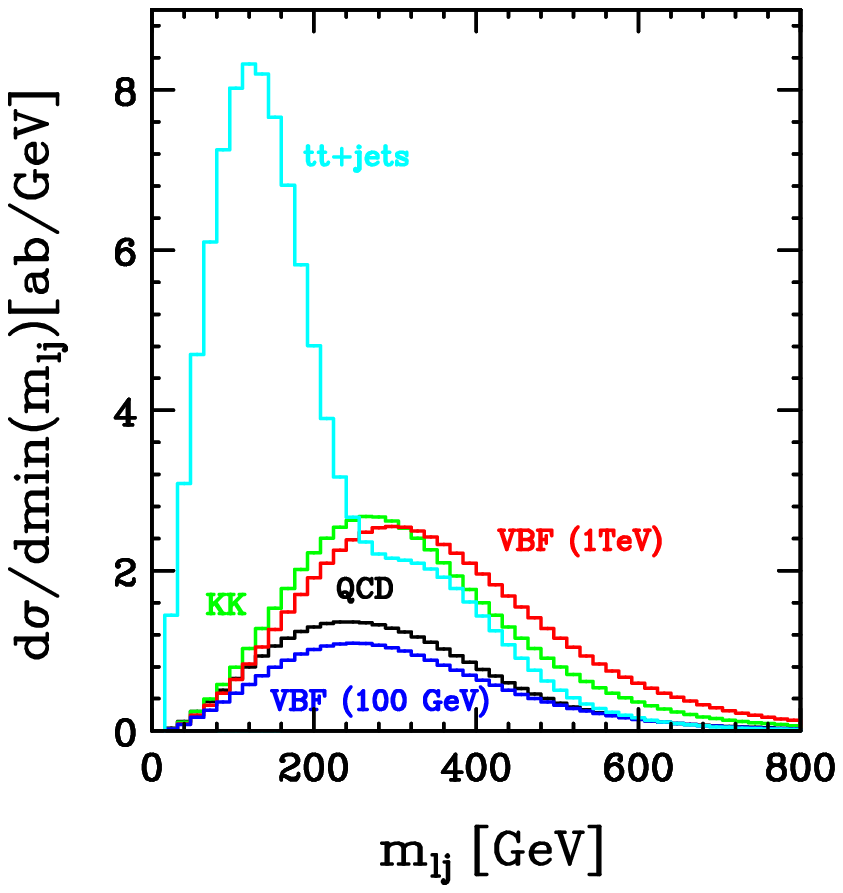}
\end{center}
\vskip -0.4cm
\ccaption{}{\label{ww3-inter} \it
Minimum invariant mass distribution of a tagging jet and a charged lepton
for the $pp \rightarrow W^+W^- jj$ process after imposing 
the cuts of Eqs.~(\ref{cuts:inc1})-(\ref{cuts:vbf4}), 
a $b$-veto, a CJV, 
and requiring $p_{T}(\ell) > 100$~GeV. Plotted are results for the 
heavy Higgs boson scenario, the Higgsless Kaluza-Klein model, and the
relevant SM backgrounds.}
\vskip 0.4cm
\end{figure}
%
is particularly effective in reducing the $t\bar{t}$ and $t\bar{t}j$ 
backgrounds, i.e. the cases where at least one of the tagging jets 
arises from a $b$-quark which is a top decay product. For top quarks which are 
almost on mass-shell, this $b$-quark must have an invariant mass with the 
charged lepton from the same top-quark decay of $m_{\ell j}<m_t$. The
$\min\,(m_{\ell j})$ cut will thus reduce the $t\bar t+$~jets background 
to mostly its $t\bar tjj$ component. At the same time, the signal processes 
are only slightly affected, as their $m_{\ell j}$ shapes are peaking 
well above 200~GeV.

The powerful sets of selection cuts introduced so far exploit the 
characteristic features of VBF processes and the fact that we are looking
for the decay products of massive objects or, more precisely, for low 
partial waves in high energy vector boson scattering. We did not impose further
leptonic cuts for the $ZZjj$ and $W^{\pm}Zjj$ channels, because the amount of 
improvement in the significance of the signal would be marginal. However, in
the case of $pp\to W^+W^-jj$ additional measures are necessary to suppress 
the overwhelming $t\bar{t} + \mr{jets}$ backgrounds. 

\item[VI.]\textsc{Central jet veto:}
QCD-induced processes tend to exhibit more jet activity in the 
central rapidity region than VBF reactions with colorless weak boson exchange
in the $t$-channel. 
A central jet veto (CJV) can therefore be applied to 
reduce QCD backgrounds by eliminating events 
where in addition to the tagging jets at high rapidity 
secondary jets with a high 
transverse momentum are found in the central regions of the detector. 

We veto any such activity by discarding all events with an 
extra veto jet of  
\beq
p_{Tj}^{veto} > 25~\mr{GeV}\,,
\eeq  
located in the gap region between the two tagging jets, 
\beq
\eta_{j,min}^{tag} < \eta_j^{veto} < \eta_{j,max}^{tag}\,.
\eeq 
In our simulations we do not yet model extra QCD radiation which might be
subject to the central jet veto. Such refinements are beyond the scope of the 
present work. However, the $t\bar{t}j$ and $t\bar{t}jj$ 
background processes typically have additional $b$-quark jets from top-quark
decay in the central region. The CJV thus is very effective in reducing these
backgrounds.
%
\begin{table}
\begin{center}
\begin{tabular}{c|c|c}
\hline
& &\\
$p_{Tj}^{veto}$ [GeV] & $1.4 < |\eta_j^{veto}| < 2.4$  & $|\eta_j^{veto}|<1.4$  
  \\
& &\\
\hline
& & \\
30 - 50   & $60\%$ & $70\%$ \\
50 - 80   & $65\%$ & $75\%$ \\
80 - 120  & $70\%$ & $80\%$ \\
120 - 170 & $70\%$ & $80\%$ \\
$>$ 170   & $65\%$ & $75\%$ \\
& &\\
\hline
\end{tabular}
\ccaption{}{\label{tabveto} \it
Assumed $b$-tagging efficiencies as functions 
of the transverse momentum of the jet for different rapidity ranges.}
\end{center}
\end{table}
%
\item[V.]\textsc{$b$-tagging jet veto:}
Discrimination between jets originating from $b$-quarks and those 
emerging from light quarks or gluons by efficient $b$-tagging 
 helps additionally to suppress 
$t\bar{t}+\mr{jets}$ backgrounds in the $W^+W^-jj$ channel: we eliminate 
any events where at least one of the tagging jets is identified as arising 
from a $b$-quark. 
We use the results of a CMS analysis~\cite{btagging} for our assumptions on
$b$-veto efficiencies and mis-tagging probabilities. For a $10\%$ mis-tagging
probability per jet one finds $b$-veto efficiencies in the range $60\%-80\%$,
depending on the transverse momentum and pseudorapidity of the jet as listed
in Tab.~\ref{tabveto}. 
\end{itemize}
%
%
%
\section{Results and Discussion}
\label{sec:num}
We now turn to a discussion of numerical results for the signal and
background processes of the scenarios discussed in the previous
sections. In all cases the cross sections correspond to two generations
of charged leptons and three neutrino species for $Z\to\bar\nu\nu$.
They are listed in Tables~\ref{tabzz}$-$\ref{tabww}
for all processes contributing to a specific leptonic final state, after
different sets of selection cuts have been applied. The impact on
inclusive cross sections of only the VBF cuts or only the leptonic cuts 
is shown in the lines labeled ``$\textsc{Inc. + Vbf}$'' 
and ``$\textsc{Inc. + Lep.}$'', respectively. In each case, we consider
the QCD $VVjj$ background, EW $VVjj$ production assuming a light or a heavy
Higgs boson, and a Warped Higgsless scenario with additional spin-one
resonances. For the $W^+W^-jj$ channel also the 
$t\bar{t} + \mr{jets}$ background is given and the impact of CJV and
$b$-veto cuts on the cross section after inclusive and VBF cuts is also
shown. In Table~\ref{tabtt}, results for
$t\bar{t}$, $t\bar{t}j$, and $t\bar{t}jj$ production are listed separately 
to better illustrate the impact of the individual contributions. 

In all channels, VBF cuts reduce the QCD $VVjj$  backgrounds efficiently,
decreasing inclusive production rates by factors of $25-85$ as shown in 
the respective second lines 
of Tables~\ref{tabzz}$-$\ref{tabww}. 
VBF cuts are even more efficient in the case
of the $t\bar{t}$, $t\bar{t}j$  and $t\bar{t}jj$  background processes, as
illustrated by Table~\ref{tabtt}. At the same time, rates for the
heavy and light Higgs boson scenarios as well as for the Higgsless 
Kaluza-Klein model have decreased by a factor of $2-3$ only.  

Imposing  leptonic cuts, on the
other hand, helps to suppress EW
backgrounds, while the respective signal processes
remain substantial, as apparent from the third rows of 
Tables~\ref{tabzz}$-$\ref{tabww}. Combining the leptonic and VBF cuts
one finds sufficient background suppression for the $ZZjj$ and $WZjj$
final states, as shown in the fourth rows of 
Tables~\ref{tabzz}$-$\ref{tabwmz}.
For the $t\bar{t}+\mr{jets}$ backgrounds to $W^+W^-jj$ final states
the impact of the leptonic cuts is
even more pronounced than  the effect of the VBF cuts. 
However, total
rates are still much higher for the backgrounds than for the  corresponding 
signal process. Thus, additional  cuts have to be applied for the $W^+W^-jj$
production mode. In order to reduce the
large $t\bar{t}+\mr{jets}$ backgrounds we make use of a $b$-veto and a CJV. 
We discard all events where one or both tagging jets can be identified
as $b$ jets, allowing for an overall
mis-tagging probability of  $10\%$ for light partons with $p_T > 30$~GeV 
and $|\eta| < 2.4$. This results in a reduction of less than $10\%$ 
for the signal and all backgrounds apart from $t\bar{t}+\mr{jets}$. 
The $b$-veto reduces these top-induced backgrounds by a factor of $2-4$.
The CJV is particularly efficient for the
$t\bar{t}jj$ process. In this case, an additional reduction factor of 18
is obtained.  The last row of 
Table~\ref{tabtt} shows that after the application of all cuts 
the $t\bar{t}+\mr{jets}$
background rates are comparable in size to those of 
the other individual backgrounds.
%
\begin{table}
\begin{center}
\begin{tabular}{c|c|c|c|c}
\hline
&&&&\\
&
QCD
&
VBF
&
VBF
&
KK
\\
Level of cuts & & $m_H =100$~GeV& $m_H =1$~TeV   & \\
&&&&\\
\hline
&&&&\\
\textsc{Inclusive}         & 3.83  & 0.2323 & 0.3101 & 0.2725 \\ 
\textsc{Inc. + Vbf}        & 0.0752 & 0.0883 & 0.1503 & 0.11152 \\ 
\textsc{Inc. + Lep.}       & 0.3755 & 0.02827 & 0.08302 & 0.04227 \\ 
\textsc{Inc. + Vbf + Lep.} & 0.00951 & 0.01171 & 0.05948 & 0.02147 \\ 
&&&&\\
\hline
\end{tabular} \ccaption{}{\label{tabzz} \it Cross sections (in fb) for various
$ZZjj \rightarrow 4\ell\,jj$ production processes  with different
Higgs boson masses and the Higgsless Kaluza-Klein scenario, after different
levels of selection cuts have  been applied, as defined in 
Section~\ref{sec:frame}. Statistical errors in all cases
are well below ~$0.5\%$. }
\end{center}
\end{table}
\begin{table}
\begin{center}
\begin{tabular}{c|c|c|c|c}
\hline
&&&&\\
&
QCD
&
VBF
&
VBF
&
KK
\\
Level of cuts &    & $m_H =100$~GeV & $m_H =1$~TeV& \\
&&&&\\
\hline
&&&&\\
\textsc{Inclusive}         & 36.13  &  1.961  &  2.482 &  2.260 \\
\textsc{Inc. + Vbf}        &  0.867  &  0.7788  &  1.196 &  0.9531 \\
\textsc{Inc. + Lep.}       &  1.717  &  0.1163  &  0.4230 &  0.1852 \\
\textsc{Inc. + Vbf + Lep.} &  0.0518 &  0.04907 &  0.3194 &  0.09883 \\
&&&&\\
\hline
\end{tabular} \ccaption{}{\label{tabzn} \it Cross sections (in fb) for various
$ZZjj \rightarrow 2\ell 2\nu\,jj$ production processes  with
different Higgs boson masses and the Higgsless Kaluza-Klein scenario, after
different levels of selection cuts have been  applied. Statistical errors in
all cases are well below $0.5\%$.}
\end{center}
\end{table}
\begin{table}
\begin{center}
\begin{tabular}{c|c|c|c|c}
\hline
&&&&\\
&
QCD
&
VBF
&
VBF
&
KK
\\
Level of cuts &    & $m_H =100$~GeV & $m_H =1$~TeV& \\
&&&&\\
\hline
&&&&\\
\textsc{Inclusive}         & 54.96  & 1.834 & 1.897 & 2.718 \\
\textsc{Inc. + Vbf}        &  2.189  & 0.6933 & 0.7382 & 1.273 \\
\textsc{Inc. + Lep.}       &  4.301  & 0.2382 & 0.2599 & 0.9161 \\
\textsc{Inc. + Vbf + Lep.} &  0.1719  & 0.0888 & 0.1077 & 0.5435 \\
&&&&\\
\hline
\end{tabular} \ccaption{}{\label{tabwpz}\it Cross sections (in fb) for various
$W^+Z jj$ production processes  with different
Higgs boson masses and the Higgsless Kaluza-Klein  scenario, after different
levels of selection cuts have been  applied. Statistical errors in all cases
are well below $0.5\%$. }
\end{center}
\end{table}
\begin{table}
\begin{center}
\begin{tabular}{c|c|c|c|c}
\hline
&&&&\\
&
QCD
&
VBF
&
VBF
&
KK
\\
Level of cuts & & $m_H =100$~GeV& $m_H =1$~TeV   & \\
&&&&\\
\hline
&&&&\\
\textsc{Inclusive}         & 37.48 & 1.1072  & 1.1445  & 1.5863 \\
\textsc{Inc. + Vbf}        & 1.304  & 0.3798  & 0.4048  & 0.6784 \\
\textsc{Inc. + Lep.}       & 2.385 & 0.1233  & 0.1344  & 0.4828 \\
\textsc{Inc. + Vbf + Lep.} & 0.0838 & 0.04324 & 0.05272 & 0.2758 \\
&&&&\\
\hline
\end{tabular} \ccaption{}{\label{tabwmz} \it Cross sections (in fb) for various
$W^-Z jj$ production processes  with
different Higgs boson masses and the Higgsless Kaluza-Klein scenario, after
different levels of selection cuts have  been applied. Statistical errors in
all cases are well below $0.5\%$.  }
\end{center}
\end{table}
\begin{table}
\begin{center}
\begin{tabular}{c|c|c|c|c|c}
\hline
&&&&&\\
&
$t\bar{t}+\mr{jets}$
&
QCD
&
VBF
&
VBF
&
KK
\\
Level of cuts &&  & $m_H =100$~GeV  &$m_H =1$~TeV & \\
&&&&&\\
\hline
&&&&&\\
\textsc{Inclusive}   & 28710   & 504.5 & 16.76 & 18.55 & 19.80 \\
\textsc{Inc. + Vbf}  &  228.667 &   5.918 &  5.063 &  6.165 & 6.536 \\
\textsc{Inc. + Lep.} &  27.4090   &   6.72 &  0.828 &  1.620 &  1.702  \\
\textsc{Inc. + Vbf} + b$-$\textsc{Veto}  
& 64.055 & 5.473 & 4.77&  5.86   & 6.22 \\
\textsc{Inc. + Vbf + CJV } &  43.197  & $-$ &  $-$ & $-$ & $-$ \\
... + b$-$\textsc{Veto}    &  24.025 &  5.47 & 4.772 & 5.856 & 6.217 \\
\textsc{... + Leptonic}    &  0.381644 &  0.202 & 0.1969 & 0.7011 & 0.588 \\
&&&&&\\
\hline
\end{tabular} \ccaption{}{\label{tabww} \it  Cross sections (in fb) 
for various
$W^+W^-jj$  production processes with different Higgs boson masses and the
Higgsless Kaluza-Klein scenario after different levels of selection cuts have
been applied. Also given is the sum of the $t\bar{t}$, $t\bar{t}j$ and 
$t\bar{t}jj$ 
backgrounds for $m_t=172.5$~GeV and $m_H=100$~GeV. Statistical errors 
are well below  $0.5\%$  for 
the $W^+W^-jj$ processes and below  $1\%$ for 
$t\bar{t}+\mr{jets}$.}
\end{center}
\end{table}
\begin{table}
\begin{center}
\begin{tabular}{c|c|c|c|c}
\hline
&&&&\\
Level of cuts
&
$t\bar{t}$
&
$t\bar{t}j$
&
$t\bar{t}jj$
&
Sum ~($t\bar{t} + \mr{jets}$)
\\
&&&&\\
\hline
&&&&\\
\textsc{Inclusive}& 13850  & 13260 & 1600  &  28710 \\
\textsc{Inc. + Vbf} & 1.967  & 131.4 & 95.3 & 228.667  \\
\textsc{Inc. + Lep.} & 0.0490 &  3.02  &  24.34 &  27.4090 \\
\textsc{Inc. + Vbf}  + b$-$\textsc{Veto} & 0.915  & 38.57 & 24.57 & 64.055 \\
\textsc{Inc. + Vbf + CJV }  &  1.967  & 35.82 & 5.41  &  43.197 \\
... + b$-$\textsc{Veto}  &  0.915 & 18.24 & 4.87 & 24.025 \\
\textsc{... + Leptonic}& 0.000844 &  0.0518 & 0.329 & 0.381644 \\
&&&&\\
\hline
\end{tabular} \ccaption{}{\label{tabtt} \it Cross sections (in fb) for the 
$t\bar{t}+ nj$ production processes, where $n=0,1,2$,  with $m_t=172.5$ 
GeV and $m_H= 100$~GeV, after different levels of 
selection cuts have been applied. Statistical errors in all cases are well
below $1\%$.}
\end{center}
\end{table}
\begin{table}
\begin{center}
\begin{tabular}{c|c|c|c|c|c|c|c}
\hline
&&&&&&&\\
Process & 
$\sigma_S$ & $\sigma_B$  & $S/B$ & $S/\sqrt{B}$ & $S/\sqrt{S+B}$ 
& $N_\mr{signal}^{SM}$ & $N_\mr{bkgd.}$
\\
&&&&&&&\\
\hline
&&&&&&&\\
$ZZ jj \rightarrow 4\ell\,jj$ & 
0.048 & 0.021 & 2.2 & 5.7 & 3.1 & 14 & 6 \\
$ZZ jj \rightarrow 2l 2\nu\, jj$ &
0.27 & 0.10 & 2.7 & 14.8 & 7.7 & 81 & 30\\
$W^+W^- jj$ & 
0.51 & 0.78 & 0.6 & 10.0 & 7.8 & 153 & 234 \\
&&&&&&&\\
\hline
&&&&&&&\\
$W^\pm Z jj$ &
0.031 & 0.386 & 0.1 & 0.9 & 0.8 & 9 & 116 \\
&&&&&&&\\
\hline
\end{tabular}
\ccaption{}{\label{tab:sb1} \it 
Cross sections for the heavy Higgs boson signal
and overall background for various channels (in fb) after all selection 
cuts have been applied. Also listed are several ratios for signal and
background rates together with the number of signal and background  
events for an assumed integrated luminosity of 300 fb$^{-1}$ at 
the LHC. }
\end{center}
\end{table}
\begin{table}
\begin{center}
\begin{tabular}{c|c|c|c|c|c|c|c}
\hline
&&&&&&&\\
Process & 
$\sigma_S$ & $\sigma_B$  & $S/B$ & $S/\sqrt{B}$ & $S/\sqrt{S+B}$  
&$N_\mr{signal}^{SM}$ & $N_\mr{bkgd.}$ \\
 &&&&&&&\\
\hline
&&&&&&&\\
$W^\pm Z jj$ &
0.68 & 0.39 & 1.7 & 18.9 & 11.4 & 204 & 117 \\
$W^+W^- jj$ & 
0.40 & 0.78 & 0.5 & 7.9 & 6.4 & 120 & 234 \\
&&&&&&&\\
\hline
&&&&&&&\\
$ZZ jj\rightarrow 4\ell\,jj$ &
0.009 & 0.021 &  0.4 & 1.1  & 0.9 & 3 & 6 \\
$ZZ jj \rightarrow 2\ell 2\nu \,jj$&
0.05 & 0.10 & 0.5  & 2.7  & 2.2 & 15 & 30 \\
&&&&&&&\\
\hline
\end{tabular}
\ccaption{}{\label{tab:sb2} \it 
Cross sections for the Higgsless Kaluza-Klein
scenario and overall background for various
channels (in fb), after  all selection cuts have been applied. Also listed are
several ratios for signal and background rates together with the 
 number of signal and 
background  events for an assumed integrated
luminosity of 300 fb$^{-1}$ at the LHC. }
\end{center}
\end{table}
In Tables~\ref{tab:sb1} and  \ref{tab:sb2}, 
the signal and combined background cross sections $\sigma_S$ and $\sigma_B$   
are listed together with the ratios $S/B$, $S/\sqrt{B}$, and $S/\sqrt{S+B}$, 
where $S$ and $B$ denote signal and background rates, respectively. 
They are calculated for a luminosity of 300 fb$^{-1}$ from the cross sections
tabulated in Tables~\ref{tabzz}--\ref{tabtt}, after all selection cuts
have been applied, with the signal defined  according to
Eqs.~(\ref{eq:signal-hh}) 
and (\ref{eq:signal-kk}). The $W^-Zjj$ channel exhibits
features very similar to the related  $W^+Zjj$ mode, while its production
rates are always smaller by approximately a factor of 2, which is due to the
size of the parton distribution functions of 
the dominant subprocesses for the respective production modes. 
In Tables~\ref{tab:sb1} and \ref{tab:sb2} we therefore combine the
$W^+Zjj$ and the $W^-Zjj$  
channels to enhance the statistical significance of the $W^\pm Z\,jj$ 
mode.

Considering the SM with a heavy Higgs boson as a prototype for scenarios
with a broad scalar, iso-scalar resonance, an indicator for the LHC
sensitivity is provided by the cross section enhancement in the 
$ZZjj$ and $W^+W^-jj$ channels for VBF with $m_H=1$ TeV. 
These two channels provide excellent possibilities for the study of  
strongly interacting gauge boson systems via scalar resonances,  see 
Table~\ref{tab:sb1}. Particularly encouraging is the signal rate for the
$ZZjj\rightarrow 2\ell 2\nu \, jj$ mode. The absence of a significant
enhancement in the $WZjj$ channel is a crucial factor in identifying the
iso-scalar character of such a resonance. 

A $5\sigma$ statistical  significance, defined here as 
$S/\sqrt{B} = 5\sigma$, for a 
signal with a heavy Higgs boson can already be obtained with 
an integrated luminosity 
of 240 fb$^{-1}$,  35 fb$^{-1}$, and 75 fb$^{-1}$, respectively, for 
the $ZZjj \rightarrow 4\ell\, jj$, the 
$ZZjj\rightarrow 2\ell 2\nu \,jj$, and the $W^+W^- jj$ processes.
It should be noted, however, that event rates for $ZZ\to 4$~charged
leptons are very small, and Poisson significances would be substantially
smaller.
$W^{\pm}Zjj$ production, with heavy Higgs boson contributions entering via 
$t$- and $u$-channel exchange diagrams only, is hardly affected by the
Higgs resonance. No significant deviation from background is expected in
this channel for the heavy Higgs scenario.

In contrast, the Warped Higgsless Kaluza-Klein model 
with a tower of additional vector resonances can be studied most easily 
via the $W^{\pm}Zjj$ and $W^+W^-jj$ modes, as shown in Table~\ref{tab:sb2}. 
In the $W^{\pm}Zjj$ channel, the first of the $W_{k}$ resonances,
$W_{2}$, can be observed. Two $Z_{k}$ resonances,  
which are difficult to disentangle,
$Z_{2}$ and $Z_{3}$, are accessible in the $W^+W^-jj$ process.  
A $5\sigma$ statistical  significance for the Higgsless signal, calculated using 
the same formula as in the heavy Higgs boson case, can be obtained
with a minimal integrated luminosity of 25 fb$^{-1}$ and 
125 fb$^{-1}$, respectively, for the $W^\pm Z jj$ and the $W^+W^- jj$
processes for our choice of the model parameter $R=9.75\times 10^{-9}$.
The two $ZZjj$ channels are much less sensitive to this model, since in
these production modes the $W_{k}$ Kaluza-Klein excitations occur only in 
$t$- and $u$-channel exchange diagrams.
A similar study for the $W^+W^-jj$ channel in the context of a Higgsless 
Kaluza-Klein scenario has been performed in Ref.~\cite{Malhotra:2006sx}, 
yielding a signal significance of comparable size.

Altogether, a reasonable number of signal events can be achieved at the LHC 
for an integrated luminosity of 300 fb$^{-1}$, see
Tables~\ref{tab:sb1}~and~\ref{tab:sb2}. Our cuts have considerably
reduced backgrounds, so that  
even a relatively small number of excess signal events should be observable. 
The $W^\pm Zjj$ channel per se is not sensitive to a scalar resonance like a 
1~TeV Higgs boson. Similarly, the $ZZjj$ mode is barely sensitive to 
the $W^\pm$ KK mode. It is however the combined analysis of all channels 
that eventually allows to select between the models as distinct realizations 
of electroweak symmetry breaking.

\begin{figure}
\begin{center}
\includegraphics[width=7.5cm,height=7.5cm]{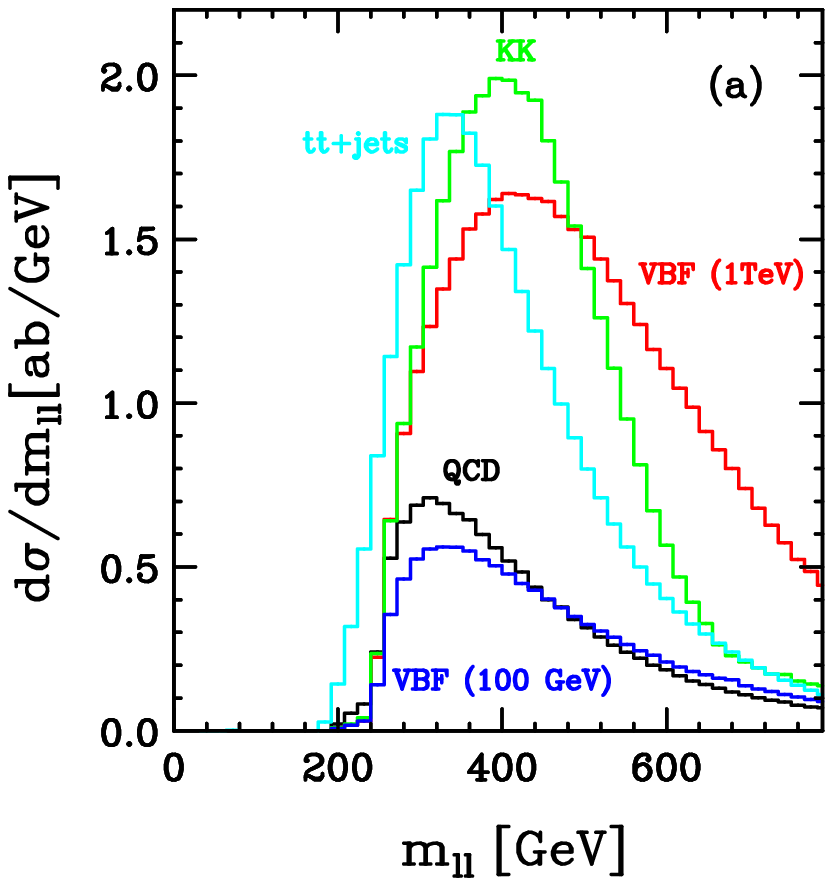}
\includegraphics[width=7.5cm,height=7.5cm]{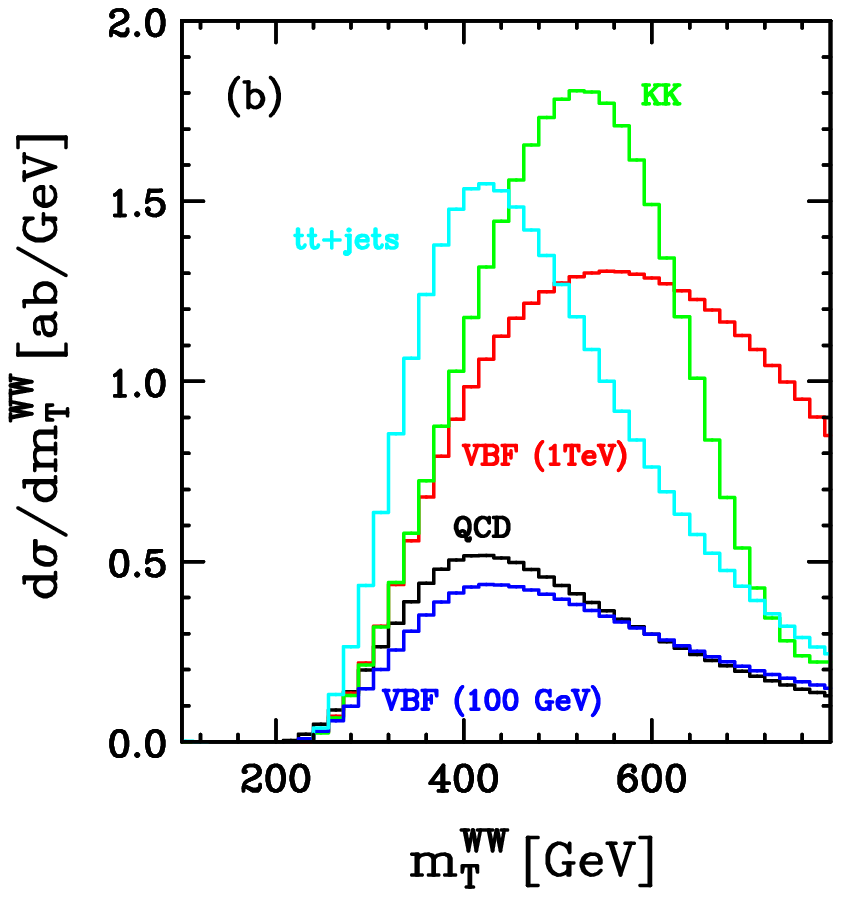}
\end{center}
\vskip -0.4cm
\ccaption{}{\label{ww1} \it
Invariant mass distribution of the  two charged leptons (a) 
and cluster transverse mass distribution of the $W^{+}W^{-}$ system (b)
for the $pp \rightarrow W^{+}W^{-} jj$ 
process after imposing all levels of cuts.}
\vskip 0.4cm
\end{figure}

In addition to the signal and background rates listed above, we have
studied various kinematic distributions for each production
process. Representative results are presented in the following, with
histograms corresponding to the cross sections listed in	
Tables~\ref{tabzz}$-$\ref{tabww}.  
Due to the large $t\bar t+\mr{jets}$ cross sections, the $W^+W^- jj$
mode  constitutes the biggest challenge.  
In Fig.~\ref{ww4-inter}~(a), we have shown the invariant mass
distribution of the two charged leptons in $pp\to W^+W^- jj$ after the
application of general selection cuts. At this level of cuts, the $t\bar
t+\mr{jets}$ background was still sizeable. If additionally all
process-specific cuts of Eq.~(\ref{eq:ww_lep_cuts}) are imposed, the
$t\bar t+\mr{jets}$ cross sections can be further reduced, while the
signal distributions are barely affected, cf.\ Fig.~\ref{ww1}~(a).  
%
%
In Fig.~\ref{ww1}~(b), the cluster transverse mass of the produced
$W^+W^-$ system, defined by  
\begin{equation}
m^{2}_{T}(WW)=
[ \sqrt{m^{2}(\ell\ell)+p_{T}^{2}(\ell\ell)} +                      
  |p_{T}^{miss}| ]^2 - [ \vec{p}_{T}(\ell\ell) +\vec{p}_{T}^{\;miss}
  ]^2\,,
\end{equation}
is shown. Similar to the $m_{\ell\ell}$ distribution, QCD and EW $VVjj$
backgrounds are small, and $t\bar t+\mr{jets}$ is well under
control. The Kaluza-Klein scenario we consider exhibits a pronounced
resonance peak, well above the backgrounds. The heavy
Higgs cross section is distributed more broadly in $m_{T}(WW)$, but
still well distinguishable.   

\begin{figure}
\begin{center}
\includegraphics[width=7.5cm,height=7.5cm]{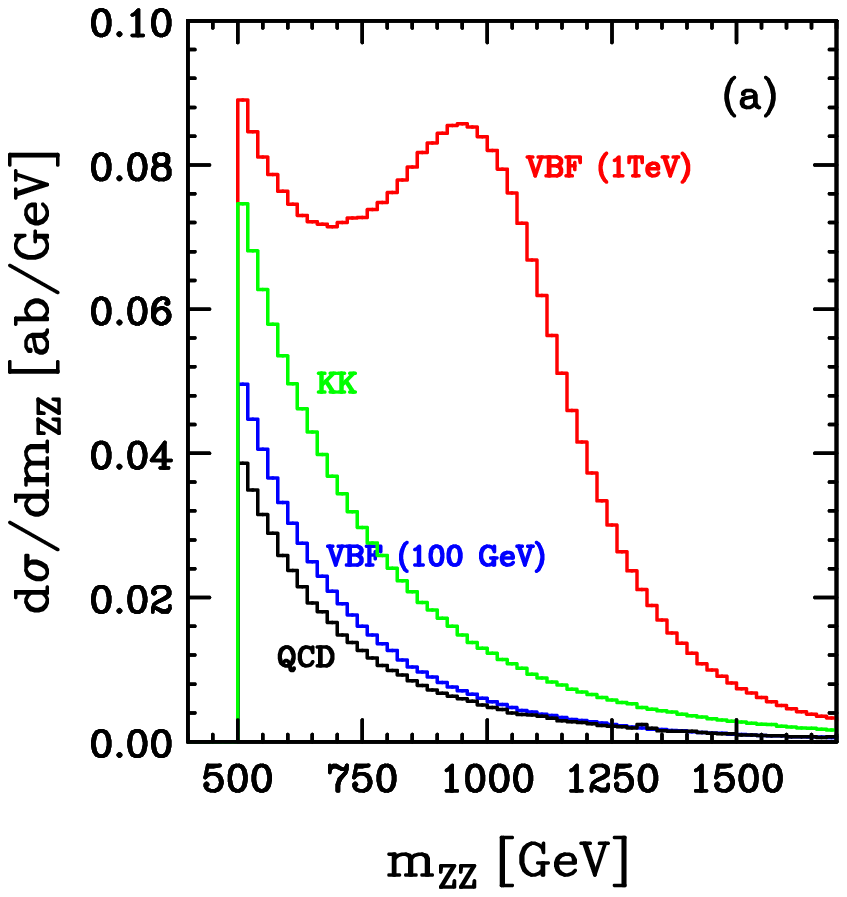}
\includegraphics[width=7.5cm,height=7.5cm]{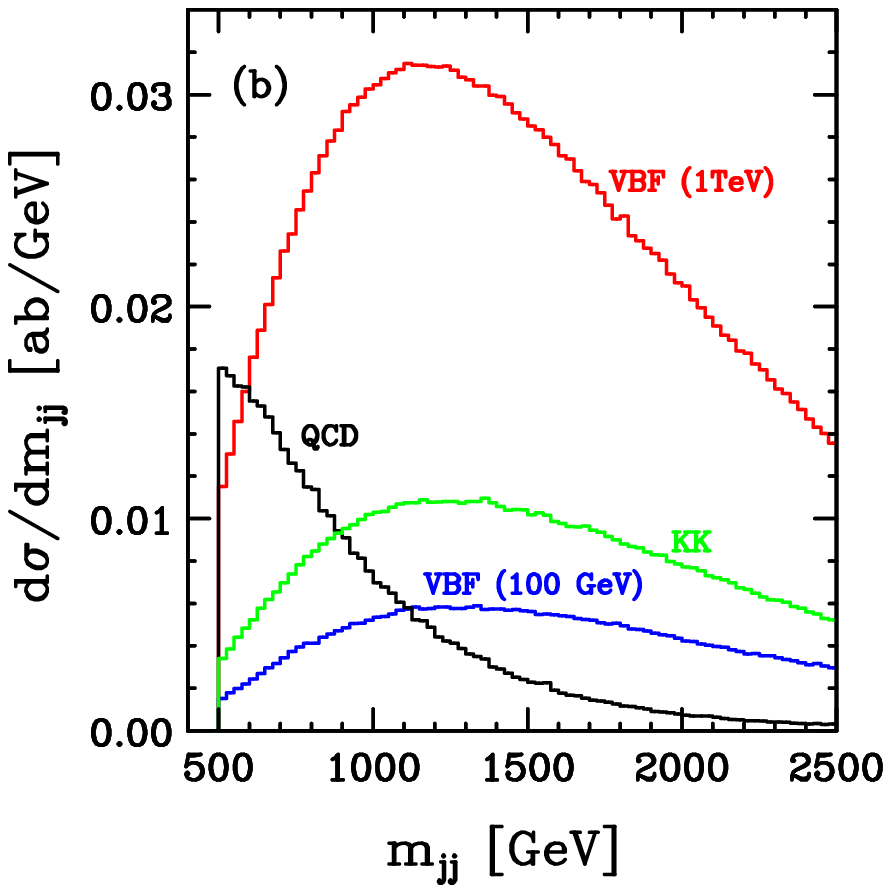}
\end{center}
\vskip -0.4cm
\ccaption{}{\label{zz1} \it
Invariant mass distribution of the four charged leptons (a) and
of the two tagging jets (b) for the 
$pp \rightarrow ZZ jj \rightarrow 4\ell\,jj $ process after
imposing all levels of cuts. }
\vskip 0.4cm
\end{figure}

The heavy Higgs scenario can also be well identified in the $ZZjj$
production modes, which are, however, less sensitive to Kaluza-Klein
resonances as discussed above.  
Figure~\ref{zz1}~(a)
%
%
shows the invariant mass distribution of the four charged leptons in $pp
\rightarrow ZZ jj \rightarrow 4\ell\,jj$ after all process-specific
selection cuts have been applied. The impact of the heavy Higgs
resonance is evident at $m_{ZZ}=1000$~GeV, where all backgrounds are
small. The Kaluza-Klein cross section exceeds the QCD and continuum EW
results, but does not exhibit a characteristic resonance behavior. 
The Higgsless model's excess over the EW continuum can be understood from the 
absence of an iso-scalar exchange contribution to weak gauge boson 
scattering, which in the SM enters with an amplitude of 
opposite phase as the gauge boson exchange graphs. 
Another
distinction can be observed in the invariant mass distribution of
the tagging jets displayed in Fig.~\ref{zz1}~(b). The excess events from
enhanced VBF production correlate with large dijet invariant masses,
while the QCD background mostly resides at $m_{jj}<1$~TeV and rapidly falls
off as $m_{jj}$ increases.  

\begin{figure}
\begin{center}
\includegraphics[width=7.5cm,height=7.5cm]{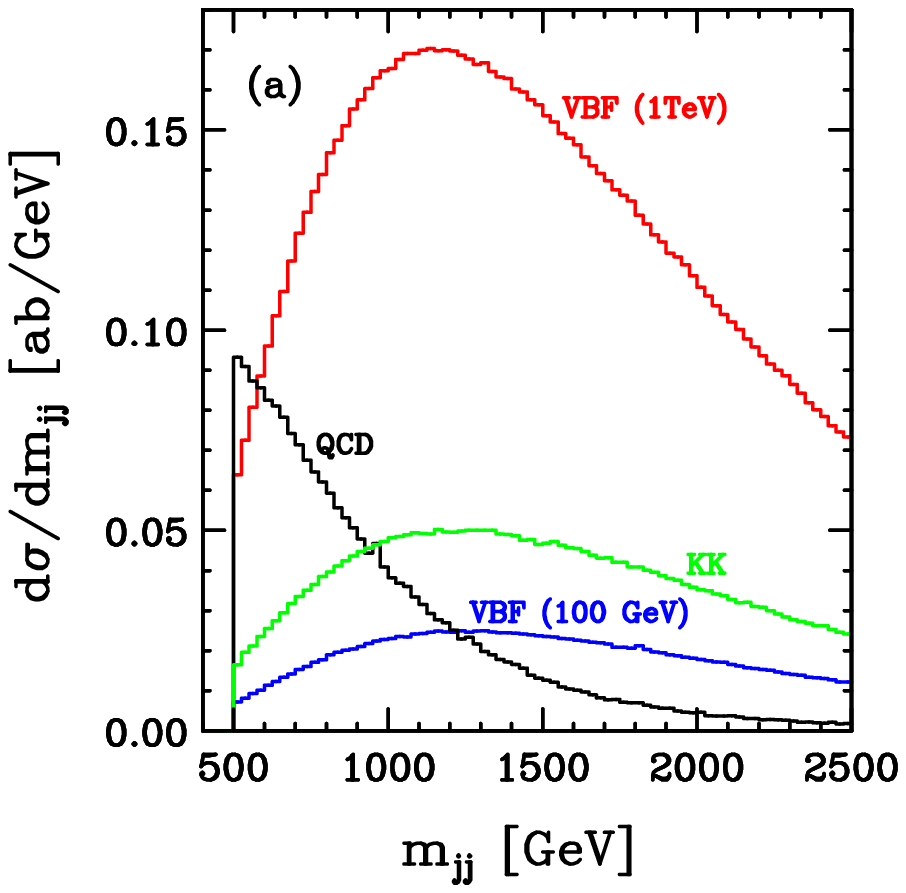}
\includegraphics[width=7.5cm,height=7.5cm]{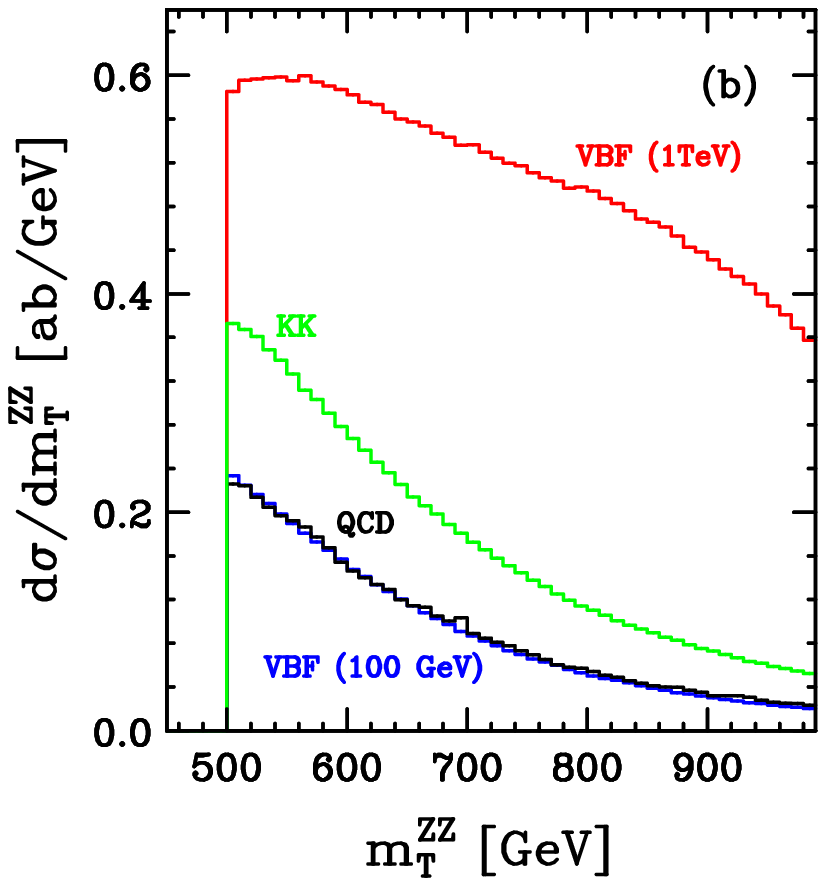}
\end{center}
\vskip -0.4cm
\ccaption{}{\label{zn1} \it
Invariant mass distribution of the two tagging jets (a)
and cluster transverse mass distribution of the $ZZ$ system (b) for the
$pp \rightarrow ZZ jj\rightarrow  2\ell 2\nu\,jj $ 
process after imposing all levels of cuts.}
\vskip 0.4cm
\end{figure}

This behavior is completely independent of the gauge boson decay, as
illustrated by Fig.~\ref{zn1}~(a), 
%
%
where the $m_{jj}$ distribution is shown for the $ZZ jj\rightarrow
2\ell 2\nu\,jj $ mode. Apparently, the shapes of the invariant mass
distribution are identical to the $ZZ jj\rightarrow  4\ell\,jj $
case. The overall normalization differs due to the $Z\to \nu\bar \nu$
branching ratio exceeding the one for $Z\to \ell^+\ell^-$.   
Fig.~\ref{zn1}~(b) illustrates the cluster transverse mass of the $ZZ$
system in the $2\ell 2\nu\,jj $ decay mode. Similar to the $m_{ZZ}$
distribution in $pp \rightarrow ZZ jj \rightarrow 4\ell\,jj$, the heavy
Higgs cross section dominates over all backgrounds. However, the Higgs
resonance does not manifest itself in a pronounced peak, but is smeared
out over a large range in $m_T(ZZ)$.  

\begin{figure}
\begin{center}
\includegraphics[width=7.5cm,height=7.5cm]{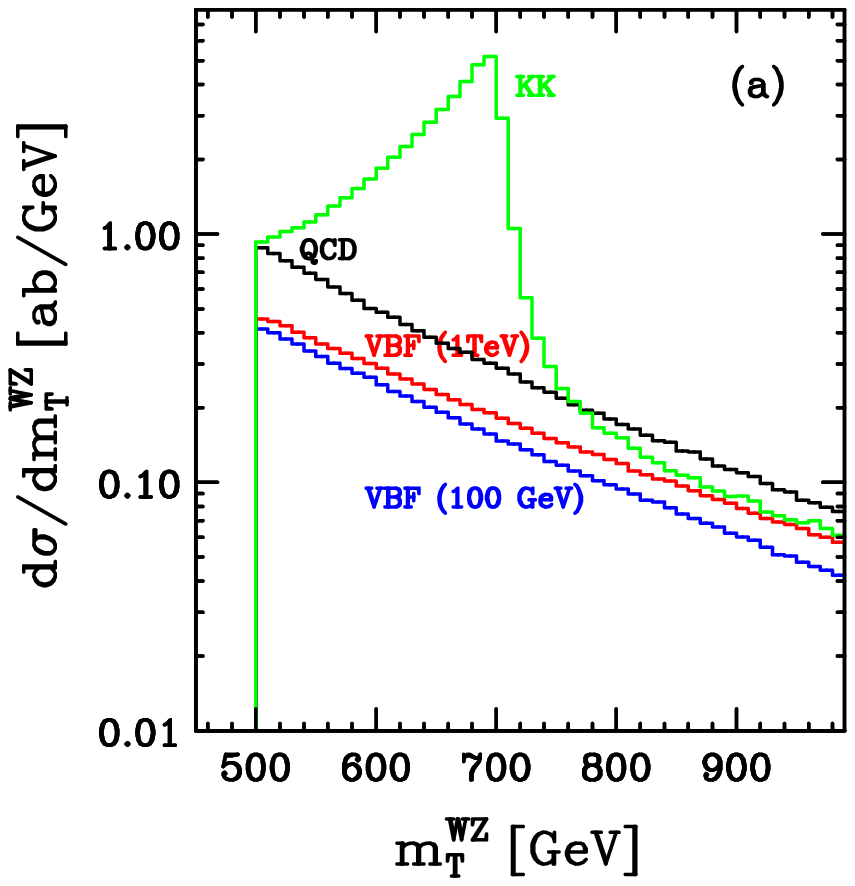}
\includegraphics[width=7.5cm,height=7.5cm]{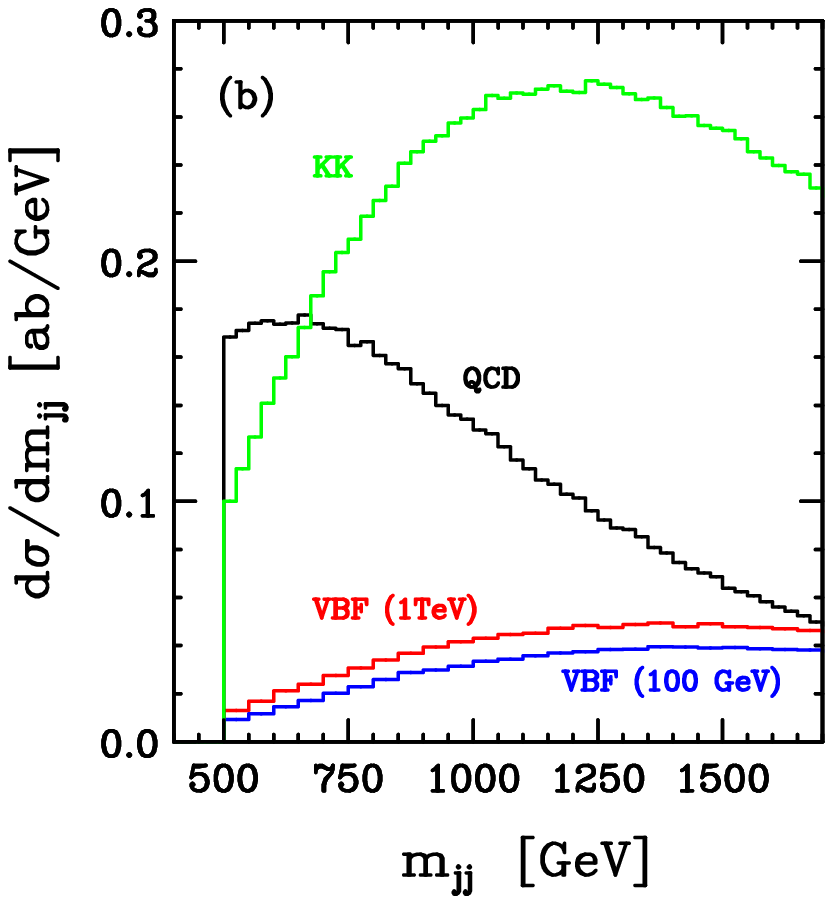}
\end{center}
\vskip -0.4cm
\ccaption{}{\label{wpz1}\it
Cluster transverse mass distribution of the $W^{+}Z$ system (a) and 
invariant mass distribution of the two tagging jets (b)
for the $pp \rightarrow W^{+}Z jj $
process after imposing all levels of cuts. }
\vskip 0.4cm
\end{figure}

The most distinctive signatures of iso-vector Kaluza-Klein excitations
are observed in the $W^\pm Zjj$ mode, since these heavy spin-one states 
contribute to resonant $W^\pm Z$
scattering, which does not occur in scenarios with a scalar Higgs boson.
This is illustrated by Fig.~\ref{wpz1}~(a), 
%
%
%
which shows the cluster transverse mass distribution for the $W^+ Zjj$
case. The $m_T(WZ)$ distribution exhibits a characteristic peak at about
$700$~GeV due to the impact of the first massive Kaluza-Klein excitation
$W_2$. The QCD and EW backgrounds as well as the heavy Higgs cross
section are smoothly distributed over $m_T(WZ)$. As expected, the VBF
cross sections for the $m_H=100$~GeV and the $m_H=1$~TeV case are very
similar, because the scalar Higgs boson contributes to $W^+Z$ scattering
only via non-resonant diagrams.  
Also for $W^+Zjj$ production, the invariant mass distribution of
the two tagging jets, as shown in Fig.~\ref{wpz1}~(b) shows the
characteristic distinction between QCD backgrounds and VBF processes. 
Shapes for $W^- Zjj$ production are almost identical to
the $W^+ Zjj$ case and therefore not displayed here.   

In summary, signatures of a heavy Higgs boson scenario as well as of
Kaluza-Klein excitations should be observable in gauge boson scattering
processes at the LHC. While an iso-scalar resonance like a heavy Higgs boson
manifests itself most distinctively in the $W^+W^-jj$ and the $ZZjj$
channels, a Higgsless scenario with iso-vector resonances, 
such as the Kaluza-Klein model which we have
considered, can be studied best in the $W^\pm Zjj$ channel.  
A multivariate analysis with full detector simulation might yield 
a slightly different shape of the distributions and total rates. 
However, our conclusions on the
observability of strong interaction signatures should remain valid. 
%
\section{Conclusions}
\label{sec:summary}
The origin of electroweak symmetry breaking is still unknown. While 
a perturbative Higgs sector with a light SM-like Higgs boson is a preferred 
solution at present, experimental tests are needed to probe other scenarios 
where new strong interactions are responsible for the weak gauge boson 
masses. With the luminosity and energy available at the LHC the search 
for such scenarios becomes feasible via studies of weak boson scattering
as realized in vector boson fusion processes. 

We have performed a broad study of 
the possibility to probe the strongly interacting
electroweak symmetry breaking sector in gauge boson scattering reactions
leading to $ZZjj$, $W^\pm Zjj$ and $W^+W^-jj$ final states at
the LHC, using only leptonic decay modes. By performing full tree-level
simulations of the dominant backgrounds, at a level where double 
forward jet-tagging acceptances can be reliably calculated, we have 
established selection cuts which allow to isolate the strong weak boson 
scattering signals. We have found that, for each
of the models we considered, an observable
excess of events occurs in at least one of the production modes, 
after three years of running with an annual luminosity
of 100 fb$^{-1}$. As compared to Ref.~\cite{Bagger:1995mk}, higher
signal rates for a SM-like 
1~TeV Higgs boson could be obtained with loosened leptonic cuts 
by tagging two jets of high transverse momenta. Defining such high 
acceptance signal regions for the various vector boson fusion channels
and providing realistic background estimates in these regions was a
second goal of our work. These results can now be used 
for further studies, be it of other scenarios for weak boson scattering,
for the assessment of higher order QCD corrections, or for refinements such as 
improved central jet veto techniques. 
We should stress that our analysis has been conservative, as a central jet 
veto offers promising prospects for a further enhancement of the 
signal-to-background ratio. From related studies on $pp\to Z+2\mr{jets}$ 
and $pp\to H+2\mr{jets}$ \cite{Rainwater:1996ud,Rainwater:1999gg} one 
expects additional suppression of the QCD backgrounds by about 70\%, while 
around 90\% of the VBF signal are retained when a central jet veto is imposed. 

In the search regions defined in Sec.~\ref{sec:frame}, ignoring improvements 
from a central jet veto on the QCD backgrounds, even the SM light Higgs scenario
yields weak boson fusion signal cross sections which are of the same order as
the QCD backgrounds, with signal to background ratios of 1:1 for $ZZjj$ final
states, 1:2 for the higher statistics $WZjj$ mode and 1:3 for $W^+W^-jj$. With
expected statistical samples of 36, 116 and 234 signal plus background events,
respectively, in 300~fb$^{-1}$ of data, an increase
of the VBF cross section by a factor of two in any of these channels should 
be observable at the LHC.  
Our conclusions are thus valid beyond the details of 
the models considered and will certainly apply to any scalar or 
vector resonances of sufficient size but general peak location.

Should a light Higgs boson be found rather than first signatures of 
strong gauge boson interactions, the precise measurement of event rates 
at high invariant mass is essential to ensure that the Higgs boson 
indeed cures the  bad high energy behavior of gauge boson scattering
processes as predicted by the SM. Given the results mentioned above, 
a measurement of the high vector boson pair invariant mass cross section
in several vector boson fusion channels, with a statistical accuracy of order
30\%, seems possible after several years of LHC running. Improvements on this
result appear feasible, in particular with a more realistic calculation of
central jet veto acceptances for QCD induced background events, which 
make use of the elevated level of central soft gluon radiation in such events.
We leave such refinements to future work.
%
%
\section*{Acknowledgments}

We are grateful to N.~Kauer for making his computer code 
for $t\bar{t}+\mr{jets}$ available for comparison with the
\textsc{Helac-Phegas} package. We furthermore 
would like to thank  G.~Kl\"amke for interesting discussions. 

Our work was supported by the Japan
Society for the Promotion of Science (JSPS) and by the Deutsche
Forschungsgemeinschaft via the Sonderforschungsbereich/Transregio SFB/TR-9
``Computational Particle Physics''. 
M.~W.\ was funded in part by the RTN European Programme 
MRTN-CT-2006-035505 HEPTOOLS - Tools and Precision Calculations 
for Physics Discoveries at Colliders, 
B.~J.\ by the Initiative and Networking Fund of the
Helmholtz Association, contract HA-101 ("Physics at the Terascale"), 
and C.~E.\  by ``CETA Strukturiertes Promotionskolleg".


\providecommand{\href}[2]{#2}\begingroup\raggedright\endgroup

\end{document}